\documentclass[prd,aps,nofootinbib,floatfix,11pt]{revtex4}
\usepackage{amsmath,graphicx,epsfig,amssymb,dsfont,mathtools}
\usepackage[usenames]{color}
\usepackage{ulem} 
\usepackage{bigstrut}
\usepackage{slashed}
\usepackage{amsthm}
\usepackage{mathrsfs}
\usepackage{multirow}
\usepackage{subfigure}
\usepackage{float}
\usepackage{braket}
\usepackage[dvipdfmx,colorlinks=true,citecolor=blue,linkcolor=blue,urlcolor=blue]{hyperref}
\allowdisplaybreaks

\begin{document}
\title{Scalar $D^{(*)}K^{(*)}$ and $D^{(*)}\pi(\rho)$ molecular states from B meson decays}
\author{Jing-Rui Shi}
\author{Jiang-Tao Zhang}
\author{Ye Xing}
\email{Corresponding author. xingye_guang@cumt.edu.cn}
\affiliation{School of Materials Science and Physics, China University of Mining and Technology, Xuzhou 221000, China}

\begin{abstract}
In this paper, we investigate the production of scalar singly charmed four-quark states from $B$ meson decays, employing phenomenological SU(3) flavor symmetry analysis and final-state interaction theory. Our study focuses on their production branching ratios and $CP$ violation.  The results show that the branching ratios for producing either $VV$ or $PP$ molecular states from $B$ meson decays can be as large as the order of $10^{-4}$. Moreover, for processes with direct $CP$ violation, the production of $VV$ molecular states exhibits a sizable $CP$ asymmetry of order $10^{-3}$. Our findings also suggest that the recently observed $T_{\bar c\bar s 0}^*(2870)^0$ state is more consistent with the $\overline D^{*0} K^{*0}$ molecular configuration. We expect that our predictions will be validated by future experiments.
\end{abstract}
\maketitle

\section{Introduction}
Since the discovery of X(3872) by the Belle experiment~\cite{Belle:2003nnu} in 2003, an increasing number of exotic hadronic states have been observed in experiments~\cite{LHCb:2021uow,LHCb:2022aki,BESIII:2020qkh,LHCb:2022ogu}, and more are expected to be discovered in the future. These states differ from conventional mesons composed of a quark-antiquark pair ($q\overline{q}$) or baryons consisting of three quarks (qqq). In 2020, the LHCb Collaboration reported the observation of two new states: scalar $T^*_{cs0}(2870)^0$($X_0(2900)$) and vector $T^*_{cs1}(2900)^0$($X_1(2900)$)~\cite{LHCb:2020pxc,LHCb:2020bls}, whose minimal quark content is identified as $cs\bar{u}\bar{d}$. This marks the first clear experimental evidence of an exotic state containing a singly charmed quark. Their masses and decay widths with statistical and systematic uncertainty are presented as,
\begin{eqnarray}
\begin{aligned}
&\text{m}(T^*_{cs0}(2870)^0)=(2866\pm 7\pm 2)\,\text{MeV},\,\Gamma(T^*_{cs0}(2870)^0)=(57\pm 12 \pm 4)\,\text{MeV},   \\
&\text{m}(T^*_{cs1}(2900)^0)=(2904\pm 5\pm 1)\,\text{MeV},\,\Gamma(T^*_{cs1}(2900)^0)=(110\pm 11 \pm 4)\,\text{MeV}.
\end{aligned}
\end{eqnarray}
In 2022, the LHCb collaboration observed another two candidates for open-charm four-quark states, scalar $T^*_{c\overline{s}0}(2900)^0$ and $T^*_{c\bar{s}0}(2900)^{++}$, in the invariant mass spectra of  $D_s^+\pi^{\pm}$ from the similar decay channels $B^0\to \overline{D}^0D_s^+\pi^-$ and $B^+\to D^-D_s^+\pi^+$  ~\cite{LHCb:2022sfr,LHCb:2022lzp}. These states are considered to possibly form the same isospin triplet, with minimal quark content $cd\bar{s}\bar{u}$ and $cu\bar{s}\bar{d}$ respectively with the same spin-parity $0^+$. Their masses and decay widths are presented as,
\begin{eqnarray}
\begin{aligned}
&\text{m}(T^*_{c\overline{s}0}(2900)^0)=(2892\pm 21\pm 2)\,\text{MeV},\,\Gamma(T^*_{c\overline{s}0}(2900)^0)=(119\pm 29)\,\text{MeV},   \\
&\text{m}(T^*_{c\overline{s}0}(2900)^{++})=2921\pm 23\pm 2)\,\text{MeV},\,\Gamma(T^*_{c\overline{s}0}(2900)^{++})=(137\pm 35)\,\text{MeV}.
\end{aligned}
\end{eqnarray}
For the singly charmed four-quark states observed so far, numerous theoretical interpretations and calculations have been proposed. These states have been studied within various frameworks, including compact tetraquark configurations~\cite{Zhang:2020oze,Wang:2020xyc,Agaev:2021knl,Ozdem:2022ydv,Wei:2022wtr,Tan:2020cpu,Liu:2022hbk,Guo:2021mja}, hadronic molecular scenarios~\cite{Xiao:2020ltm,Agaev:2020nrc,Chen:2020aos,Dong:2020rgs,Molina:2020hde}, and kinematical effects such as triangle singularities~\cite{Liu:2020orv,Burns:2020epm}. Within these assumptions, their properties have been extensively investigated using various theoretical tools, such as QCD sum rules~\cite{Zhang:2020oze,Wang:2020xyc,Agaev:2021knl,Agaev:2020nrc,Yang:2024coj,Albuquerque:2020ugi} and final-state interactions~\cite{Xiao:2020ltm,Ding:2024dif,Yu:2023avh,Wang:2023hpp,Chen:2020eyu}, aiming to deepen the study of them.

This paper proceed with the work under the premise of molecular state for the scalar singly charmed four-quark state $T_{cs(\bar{s})}$. Given that the quantum numbers $J^P=0^+$, they then can be a pseudoscalar-pseudoscalar (PP) meson pair or a vector-vector (VV) meson pair. To thoroughly investigate their possible production mechanism, we systematically study the production channels, production branching ratios, and CP violation of singly charmed four-quark states from B meson decays, by applying the SU(3) flavor symmetry analysis and final-state interaction approach. These two methods have been widely applied in the field of heavy meson systems~\cite{Cheng:2004ru,He:1998rq,Chiang:2004nm,Shi:2017dto,Xing:2018bqt,Hu:2022qlr,Li:2024fmg,Hsiao:2015iiu,Savage:1989ub,Gronau:1995hm,He:2000ys,Li:2007bh}. Based on flavor symmetry, it is straightforward to perform a simple phenomenological analysis of the production channels for further study. Additionally, the final-state interaction approach is utilized to the explicit calculation of the branching ratios and CP violation. Since they could further be $DK$ ($D^*K^*$), $D_s\pi$($D_s^*\rho$), or a mixture of $DK$ ($D^*K^*$) and $D_s^-\pi$($D_s^*\rho$) molecular states, the production branching ratios are calculated separately in the paper. In B meson decays, owing to the interference between tree and penguin diagrams together with final-state interaction effects, direct CP violation in the production processes could be sizable.

The remainder of this paper is organized as follows. In section.~\ref{sec:2}, we introduce all multiplets of the singly charmed four-quark and other hadrons employed in the SU(3) flavor symmetry. And giving the effective Hamiltonian for the production processes at the hadron level, as well as the relations for decay widths of different channels. Section.~\ref{sec:3} presents the detailed calculations based on final-state interaction approach, including short- and long-distance effects. Section.~\ref{sec:4} presents the numerical results and  discussion about the branching ratio and CP violation. Finally, the section.~\ref{sec:6} concludes this paper.

\section{Phenomenological Analysis: SU(3) Flavor symmetry}
\label{sec:2}
\begin{figure}
\centering
\includegraphics[scale=0.45]{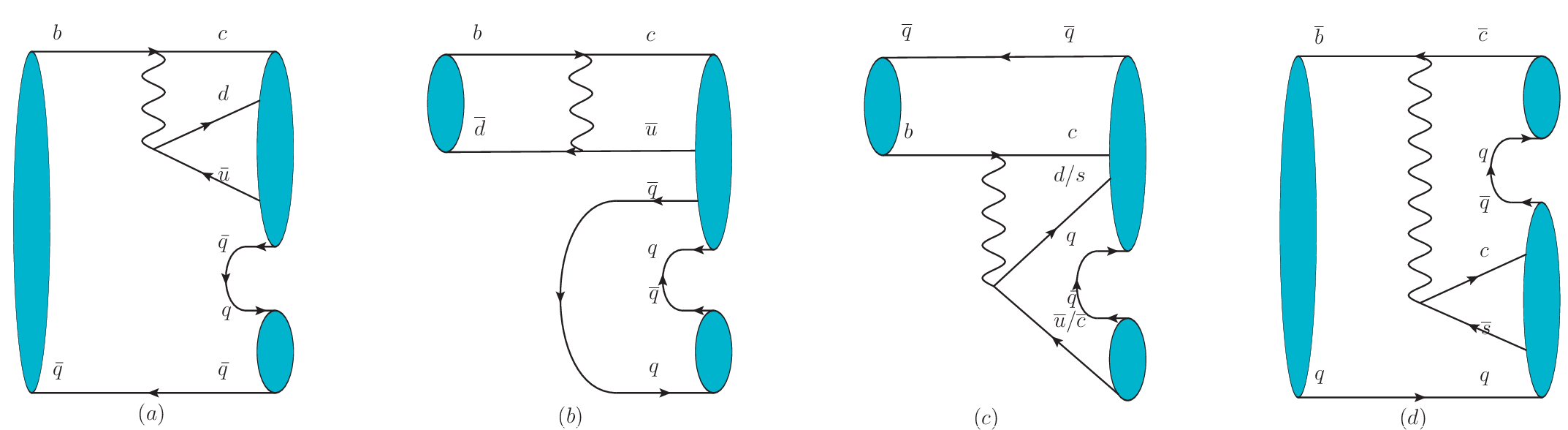}
\caption{The topological diagrams of these possible production modes. Diagram (a, d): the internal W-emission diagram; Diagram (b): the W exchange diagram; Diagram (c): the external W-emission diagram.}
\label{Fig:1}
\end{figure}

The SU(3) flavor symmetry analysis enables the study of various decay processes without relying on the specific dynamics of factorization. It has been widely used to the weak decays of heavy hadrons.  For the singly charmed four-quark states, the flavor multiplets are decomposed into anti-triplet, sextet and 15-plet. Here we only force on the sextet~\cite{Xing:2019hjg}, especially the states with different quark flavor constituents, $T_{cu\bar{d}\bar{s}}$, $T_{cd\bar{s}\bar{u}}$, and $T_{cs\bar{u}\bar{d}}$. The multiplet can be written as~\cite{He:2016yhd},
\begin{eqnarray}
&&(T_{c6})^{1}_{[23]}=\frac{1}{\sqrt{2}}T_{cu\bar{d}\bar{s}},\,(T_{c6})^{2}_{[31]}=\frac{1}{\sqrt{2}} T_{cd\bar{s}\bar{u}},\,(T_{c6})^{3}_{[12]}=\frac{1}{\sqrt{2}}T_{cs\bar{u}\bar{d}},\notag\\
&&(T_{c6})^{1}_{[12]}=(T_{c6})^{3}_{[23]}=\frac{1}{2}T_{c(u\bar{u},s\bar{s})\bar{d}},\, (T_{c6})^{1}_{[31]}=(T_{c6})^{2}_{[23]}=\frac{1}{2}T_{c(u\bar{u},d\bar{d})\bar{s}},\notag\\
&&(T_{c6})^{2}_{[12]}=(T_{c6})^{3}_{[31]}=\frac{1}{2}T_{c(d\bar{d},s\bar{s})\bar{u}}.
\end{eqnarray}
Initial bottom meson can form an triplet, while the light pseudoscalar meson ($P$) and light vector meson ($V$) will both form an octet, the octet multiplets are shown as,
\begin{eqnarray}
P_{8} = \begin{pmatrix}
 \frac{\pi^0}{\sqrt{2}}+\frac{\eta}{\sqrt{6}}  &\pi^+ & K^+\\
 \pi^-&-\frac{\pi^0}{\sqrt{2}}+\frac{\eta}{\sqrt{6}}&{K^0}\\
 K^-&\overline K^0 &-2\frac{\eta}{\sqrt{6}}
 \end{pmatrix},
&
V_{8} = \begin{pmatrix}
 \frac{\rho^0}{\sqrt{2}}+\frac{\omega}{\sqrt{2}}  &\rho^+ & K^{*+}\\
 \rho^-&-\frac{\rho^0}{\sqrt{2}}+\frac{\omega}{\sqrt{2}}&{K^{*0}}\\
 K^{*-}&\overline K^{*0} & \phi
 \end{pmatrix}.
\end{eqnarray}
The singly charmed four-quark state can be produced by the transition of $b \rightarrow c\bar{u}d/s$, $b \rightarrow c\bar{c}s/d$ and $b\rightarrow u \bar ud$, whose operators can form SU(3) irreducible representation operators $H_8$, $H_3$, and ($H_3^t$, $H_{\bar 6}^t$, $H_{15}^t$) respectively. The non-zero components are $(H_8)^{2}_{1}=V_{cb}V_{ud}^*$, $(H_8)^{3}_{1}=V_{cb}V_{us}^*$, $(H_3)^2=V_{cb}V_{cd}^*$, $(H_3)^3=V_{cb}V_{cs}^*$, and $(H_{3}^t)^2=V_{ub}V_{ud}^*$, $(H_{\bar 6}^t)_{\{13\}}=V_{ub}V_{ud}^*$, $(H_{15}^t)^{\{12\}}_1=3 V_{ub}V_{ud}^*$, $(H_{15}^t)^{22}_2=-2 V_{ub}V_{ud}^*$, $(H_{15}^t)^{\{23\}}_{3}=-V_{ub}V_{ud}^*$ ~\cite{Xing:2019hjg}. With the multiplet and operator, we can construct the possible effective Hamiltonian for the production at the hadron level,
\begin{eqnarray}
\mathcal{H}_{T_c}^{q\bar q}&=&a_{1}B_{j}(H_8)^{i}_{l}M^{j}_{k}(T_{c6})^{[kl]}_{i}+a_{2}B_{l}(H_8)^{i}_{j}M^{j}_{k}(T_{c6})^{[kl]}_{i}+a_{3}B_{l}(H_8)^{i}_{k}M^{j}_{i}(T_{c6})^{[kl]}_{j} \notag \\
   &&+a_{4}B_{i}(H_8)^{i}_{k}M^{j}_{l}(T_{c6})^{[kl]}_{j}+b_{1}B_{k}(H_3)^{i}\overline{D}_{j}(T_{c6})^{[kj]}_{i}+c_{1}\overline{B}^{i}(H_{\overline{3}})_{k}\overline{D}_{j}(T_{c})^{[kj]}_{i},\\
\mathcal{H}_{T_c}^{c\bar c}&=&d_1 B_{k} (H^t_3)^i \overline {D}_{j} (T_{c6})^{[kj]}_{i}+d_2 B_k (H^t_{\bar 6})_{\{il\}} \overline{D}^{[ij]} (T_{c6})^{[kl]}_{j}+d_3 B_i (H^t_{15})^{\{ij\}}_{k} \overline{D}_{l} (T_{c6})^{[kl]}_{j}\notag\\
   &&+d_4 B_k (H^t_{15})^{\{ij\}}_l \overline{D}_j (T_{c6})^{[kl]}_i+d_1'\overline {B}^i(H^t_{\bar 3})_k \overline D_j (T_{c6})^{[kj]}_{i}+d_2' \overline {B}_{[ik]}(H^t_{6})^{\{ij\}} \overline D_l (T_{c6})^{[kl]}_{j}\notag\\
   &&+d_3' \overline {B}_{[kl]}(H^t_{6})^{\{ij\}} \overline D_i (T_{c6})^{[kl]}_{j}+d_4' \overline {B}^i(H^t_{15})^j_{\{ik\}} \overline D_l (T_{c6})^{[kl]}_{j}.
\end{eqnarray}
Here, $\mathcal{H}_{T_c}^{q\bar q}$ and $\mathcal{H}_{T_c}^{c\bar c}$ are associated with transition $b\to c$ and $b\to u$, respectively. The coefficients $a_i$, $b_1$, $c_1$, $d_i$, and $d_i'$ represent the non-perturbative effect. The topological diagrams of these possible production modes from Hamiltonian $\mathcal{H}_{T_c}^{q\bar q}$ are shown in Fig~\ref{Fig:1}. The $a_1$, $a_4$, and $c_1$ terms correspond to Figs.~\ref{Fig:1}(a), (b), and (d), respectively, and the $a_2$, $a_3$, $b_1$ terms correspond to Fig.~\ref{Fig:1}(c).
The $\mathcal{H}_{T_c}^{q\bar q}$ is related to the process of producing light quark pairs $q\bar q$, while $\mathcal{H}_{T_c}^{c\bar c}$ requires the creation of a charm quark pair to form the corresponding final-state hadrons. Expanding the effective Hamiltonian yields the possible production processes. The corresponding channels and amplitudes are listed in Tab.~\ref{tab:3}. It can be readily seen from the table that the operators of $b\to c\bar c q$ and $b\to u\bar u q$ can both contribute to the same process. When neglecting the effects of phase-space and small CKM matrix elements $V_{ub}^*V_{us}$, we can further derive relations between different channels.
\begin{eqnarray}
&&\Gamma(B^- \to \pi^- T_{cs\bar{u}\bar{d}}^{0})=2\Gamma(\overline B^0\to \pi^0 T_{cs\bar{u}\bar{d}}),\notag\\
&&\Gamma(B^- \to D^- T_{cs\bar{u}\bar{d}}^{0})
\approx\Gamma(\overline B^0 \to \overline D^0 T_{cs\bar{u}\bar{d}}^{0}),\ \Gamma(B^+\to   D^-  T_{cu\bar{d}\bar{s}}^{++})
\approx\Gamma(B^0\to   \overline D^0  T_{cd\bar{s}\bar{u}}^{0}).\label{eq:CArelation}
\end{eqnarray}
According to the CKM matrix element in amplitudes, we then simply define ratio $R_1^{SU(3)}$ and $R_2^{SU(3)}$ to relate the Cabibbo-allowed(CA) and Cabibbo-suppressed(CS) processes,
\begin{eqnarray}
&&R_1^{SU(3)}=\frac{\Gamma(\overline B^0\to\overline{K}^0 T_{cd\bar s\bar u}^0)}{\Gamma(\overline B_s^0\to K^0 T_{cs\bar u\bar d}^0)}=\frac{\Gamma(\overline B^0\to K^0 T_{cs\bar u\bar d}^0)}{\Gamma(\overline B_s^0\to \overline{K}^0 T_{cd\bar s\bar u}^0)}=\frac{\Gamma(B^- \to K^- T_{cd\bar{s}\bar{u}}^{0})}{\Gamma(B^- \to \pi^- T_{cs\bar u\bar d}^{0})}=\Big|\frac{V_{ud}}{V_{us}}\Big|^2,\notag\\
&&R_2^{SU(3)}=\frac{\Gamma(B^-\to D_s^- T_{cd\bar s\bar u}^0)}{\Gamma(\overline B^0\to \overline{D}^0 T_{cs\bar u\bar d}^0)}=\frac{\Gamma(B_s^0\to \overline{D}^0 T_{cs\bar u\bar d}^0)}{\Gamma(B^+\to D^- T_{cu\bar d\bar s}^{++})}\sim\Big|\frac{V_{cd}}{V_{cs}}\Big|^2.\label{eq:CSrelation}
\end{eqnarray}
Noted that all the results of SU(3) analysis does not depend on the inner structure of the four-quark state, regardless of whether it is a compact state, $D^{(*)}K^{(*)}$ molecular states , or $D^{(*)} \pi (\rho)$ molecular states.
\begin{table}
\caption{The possible production processes and amplitudes of singly charmed four-quark state $T_{c6}$.}
\label{tab:3}
\centering
\begin{tabular}{|cc|cc|}\hline
channel & amplitude & channel & amplitude \\\hline
$B^-\!\!\to   \pi^-   T_{cs\bar{u}\bar{d}}^{0} $  & $ \frac{-\left(a_1+a_2\right) V_{cb}V_{us}^*}{\sqrt{2}}$ &
$\overline{B}^0_s\!\!\to   \overline D^0  T_{cd\bar{s}\bar{u}}^{0} $ & $ \frac{b_1 V_{cb}V_{cd}^*+(d_1-d_2-d_3+3d_4)V_{ub}V_{ud}^*}{\sqrt{2}}$ \\
$B^-\!\!\to   K^-   T_{cd\bar{s}\bar{u}}^{0} $ & $ \frac{\left(a_1+a_2\right) V_{cb}V_{ud}^*}{\sqrt{2}}$ &
$B^-\!\!\to   D^-  T_{cs\bar{u}\bar{d}}^{0} $ & $ \frac{b_1 V_{cb}V_{cs}^*+(d_1+d_2+3d_3-d_4)V_{ub}V_{us}^*}{\sqrt{2}}$ \\
$\overline{B}^0\!\!\to   \pi^0   T_{cs\bar{u}\bar{d}}^{0} $ & $ \frac{\left(a_1+a_2\right)V_{cb}V_{us}^*}{2} $  &
$\overline{B}^0\!\!\to  \overline D^0  T_{cs\bar{u}\bar{d}}^{0} $ & $ \frac{(d_2+d_3-d_1-3d_4)V_{ub}V_{us}^*-b_1 V_{cb}V_{cs}^*}{\sqrt{2}}$ \\
$\overline{B}^0\!\!\to   K^0   T_{cs\bar{u}\bar{d}}^{0} $ & $ \frac{\left(a_3+a_4\right) V_{cb}V_{ud}^*}{\sqrt{2}}$    &
$B^-\!\!\to    D^-_s  T_{cd\bar{s}\bar{u}}^{0} $ & $ \frac{(d_4-d_1-d_2-3d_3)V_{ub}V_{ud}^*-b_1 V_{cb}V_{cd}^*}{\sqrt{2}}$ \\
$\overline{B}^0\!\!\to   \overline K^0   T_{cd\bar{s}\bar{u}}^{0} $ & $ \frac{\left(a_1-a_4\right) V_{cb}V_{ud}^*}{\sqrt{2}}$  &
$B^0\!\!\to   \overline D^0  T_{cd\bar{s}\bar{u}}^{0} $ & $ \frac{c_1 V_{cb}^* V_{cs}+(d_1'+d_2'-2d_3'-d_4')V_{ub}^*V_{us}}{\sqrt{2}}$ \\
$\overline{B}^0_s\!\!\to   \pi^0   T_{cd\bar{s}\bar{u}}^{0} $ & $ \frac{\left(a_3-a_2\right) V_{cb}V_{ud}^*}{2}$ &
$B^+\!\!\to    D^-_s  T_{cu\bar{d}\bar{s}}^{++} $ & $ \frac{c_1 V_{cb}^*V_{cd}+(d_1'-d_2'+2d_3'+3d_4')V_{ub}^*V_{ud}}{\sqrt{2}}$ \\
$\overline{B}^0_s\!\!\to   K^0     T_{cs\bar{u}\bar{d}}^{0} $ & $ \frac{\left(a_4-a_1\right) V_{cb}V_{us}^*}{\sqrt{2}}$   &
$B^0_s\!\!\to   \overline D^0  T_{cs\bar{u}\bar{d}}^{0} $ & $ \frac{(2d_3'+d_4'-d_1'-d_2')V_{ub}^*V_{ud}-c_1 V_{cb}^*V_{cd}}{\sqrt{2}}$ \\
$\overline{B}^0_s\!\!\to   \overline K^0   T_{cd\bar{s}\bar{u}}^{0} $ & $ \frac{-\left(a_3+a_4\right) V_{cb}V_{us}^*}{\sqrt{2}}$ &
$B^+\!\!\to   D^-  T_{cu\bar{d}\bar{s}}^{++} $ & $ \frac{(2d_3'+3d_4'-d_1'-d_2')V_{ub}^*V_{us}-c_1 V_{cb}^*V_{cs}}{\sqrt{2}}$ \\
\hline
\end{tabular}
\end{table}
\section{The Analysis by the Final states interaction Approach}
\label{sec:3}
\begin{figure}
\centering
\includegraphics[scale=0.65]{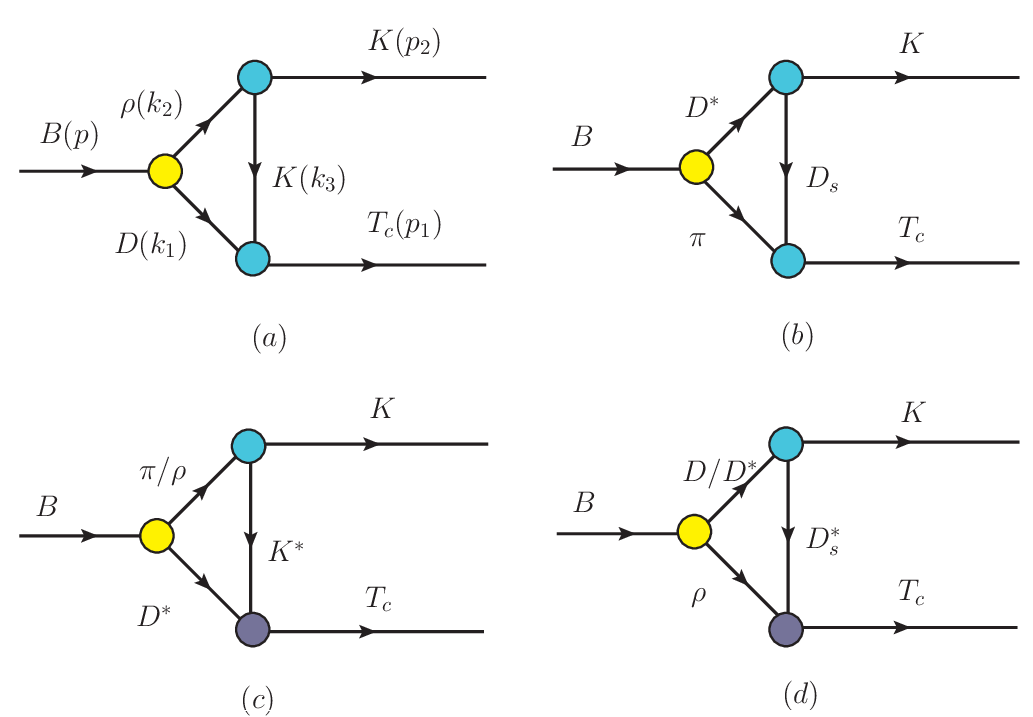}
\caption{The triangle diagrams relevant to the process $B^-\to   K^-   T_{cd\bar{s}\bar{u}}^{0}$. Diagram (a) and (b) illustrate the production mechanisms of the $T_{cd\bar{s}\bar{u}}^{0}$ in its PP molecular configuration, whereas diagram (c) and (d) depict the corresponding production processes for the VV molecular configuration of $T_{cd\bar{s}\bar{u}}^{0}$.}
\label{Fig:2}
\end{figure}
Under the molecular state hypothesis, singly charmed four-quark states with $J^P=0^+$ may possess either PP meson or VV meson configurations. For the three states of interest, such as $T_{cu\bar d\bar s}$, within the PP(VV) meson configuration, they could further be $DK$ ($D^*K^*$) molecular states, $D_s\pi$($D_s^*\rho$) molecule, or a mixture of $DK$ ($D^*K^*$) and $D_s^-\pi$($D_s^*\rho$)~\cite{Xiao:2020ltm,Agaev:2020nrc,Chen:2020aos}.  These three cases are discussed separately. In the final-state interaction approach, the production of molecular states from B meson consists of short- and long-distance processes.  The short-distance process, corresponding to the weak decay of the B meson, is characterized by form factors and decay constants. The long-distance process, arising from the strong interaction of the final hadrons from the weak decay, is described by the hadronic effective Lagrangian. For instance, the possible triangle diagrams for the $B^-\to   K^-   T_{cd\bar{s}\bar{u}}^{0}$ process are depicted in Fig ~\ref{Fig:2}.

The short distance can be described by weak effective Hamiltonian,
\begin{eqnarray}
\begin{aligned}
\label{Ham}
\mathcal{H}_{eff}=\frac{G_F}{\sqrt{2}}\left(V_{qb}V_{q_iq_j}^*(C_1 O_1+C_2 O_2)-V_{tb}V_{tq_j}^*\sum_{i=3}^{10}C_{i}O_i \right),
\end{aligned}
\end{eqnarray}
$C_i$ and  $V_{ij}$ denote the Wilson short-distance coefficient and  CKM matrix element. $O_i$s are four-fermion operators. Based on the naive factorization scheme, the weak process can be expressed as
\begin{eqnarray}
\label{FM}
&\mathcal{M}(B\rightarrow M_1 M_2) &= \frac{G_F}{\sqrt{2}}V_{qb}V_{q_iq_j}^* \Big( a_{1} \bra{M_1}(\bar{q}b)_{V\!-\!A}\ket{B}\bra{M_2}(\bar{q_i}q_j)_{V-A}\ket{0} \notag\\
& &+a_{2} \bra{M_2}(\bar{q_i}b)_{V-A}\ket{B}\bra{M_1}(\bar{q}q_j)_{V-A}\ket{0}\Big)\notag\\
&&-\frac{G_F}{\sqrt{2}}V_{tb}V_{tq_j}^*\Big(a_3' \bra{M_1}(\bar{q_j}b)_{V-A}\ket{B}\bra{M_2}(\bar{q_i}q_i)_{V-A}\ket{0}\notag\\
&&+a_5' \bra{M_1}(\bar{q_j}b)_{V-A}\ket{B}\bra{M_2}(\bar{q_i}q_i)_{V+A}\ket{0}\Big),
\end{eqnarray}
here, $M_i$ represents pseudoscalar ($P$) or vector ($V$) meson. Only the tree operator ($O_1, O_2$) contribute to the transition $b \rightarrow c\bar{u}d$, whereas the transition $b \rightarrow c\bar{c}s$ and $b\rightarrow u \bar ud$ receive contribution from penguin operators. The effective Wilson coefficients are defined as
\begin{equation}
\begin{aligned}
a_{1}=&C_{1}+\frac{C_2}{N_c},\\
a_{2}=&C_{2}+\frac{C_1}{N_c},\\
a_3'=&C_3+\frac{C_4}{N_c}+ \frac{9}{4} \left(C_9+\frac{C_{10}}{N_c}\right) , \\
a_5'=&C_5+\frac{C_6}{N_c}+ \frac{9}{4} \left(C_7+\frac{C_8}{N_c}\right)  ,  \\
\end{aligned}
\end{equation}
where $N_{c}$ represents the number of colors.  The form factors are defined as follows~\cite{McLean:2019qcx,Harrison:2021tol}:
\begin{eqnarray}
\left\langle P\left(k_1\right)\right|(\overline{q} b)_{V-A}|B(p)\rangle=&F_1(k_2^2)\Big(p_{\mu}+k_{1\mu}-\frac{m_B^2-m_P^2}{k_2^2} k_{2\mu}\Big)+F_2(k_2^2)\Big(\frac{m_B^2-m_P^2}{k_2^2} k_{2\mu}\Big), \\
\left\langle V\left(k_1\right)\right|(\overline{q} b)_{V-A}|B(p)\rangle =&\Big(\frac{2 i }{m_B+m_{V}} \varepsilon_{\mu \nu \rho \sigma} \epsilon^{*\nu}(k_1) k_{1}^{ \rho} p^\sigma  \Big) V(k_2^2)
- \Big( 2 m_{V}  \frac{\varepsilon^*(k_1) \cdot k_2}{k_2^2} k_{2\mu} \Big) A_0(k_2^2)  \notag\\
&-(m_B+m_{V})\Big(\varepsilon^{*}_{ \mu}(k_1)-\frac{\varepsilon^*(k_1) \cdot k_2}{k_2^2} k_{2\mu}\Big)  A_1(k_2^2)  \notag\\
&+ \frac{\varepsilon^*(k_1) \cdot k_2}{m_B+m_{V}}\Big(p_\mu+k_{1\mu}-\frac{m_B^2-m_{V}^2}{k_2^2} k_{2\mu}\Big) A_2(k_2^2).
\end{eqnarray}
Here transferred momentum $k_2=p-k_1$, $\epsilon^{*\nu}(k_1)$ represents the polarization vector of vector meson. In addition, the decay constants are defined as~\cite{Cheng:1996cs}:
\begin{eqnarray}
\langle P(k_1) |(\bar{q} q)_{V-A}|0\rangle=i f_P k_1^\mu, \quad\langle V (k_1) |(\bar{q} q)_{V-A}|0\rangle=m_V f_V \varepsilon^{* \mu}(k_1) .
\end{eqnarray}
The effective hadronic Lagrangian~\cite{Shen:2019evi} describing the long-distance strong interaction between hadrons, where the forms are
\begin{eqnarray}
&\mathcal{L}_{D D J/\psi}= & i g_{\scalebox{0.5}{$DD J\psi$}}J/\psi_\alpha(\partial^\alpha D \overline{D}-D\partial^\alpha \overline{D}),  \notag\\
&\mathcal{L}_{D^* D J/\psi}= & - g_{\scalebox{0.5}{$D^*D J\psi$}} \varepsilon^{\alpha \beta \rho \tau}\partial_\alpha J/\psi_\beta (\partial_\rho D^*_{\tau} \overline{D}+D\partial_\rho \overline{D}^*_\tau),\notag\\
&\mathcal{L}_{D D^* \eta_c}= & i g_{\scalebox{0.5}{$D D^* \eta_c$}}\left(D^{*\alpha}(\partial_\alpha \eta_c \overline{D}-\eta_c\partial_\alpha \overline{D}) +  (\partial_\alpha \eta_c D-\eta_c\partial_\alpha D)\overline{D}^{*\alpha}     \right), \notag\\
&\mathcal{L}_{\mathcal{V} \mathcal{P}_1 \mathcal{P}_2}= &i g_{\scalebox{0.5}{$\mathcal{V} \mathcal{P}_1 \mathcal{P}_2$}}\left(\mathcal{V}_\mu \partial^\mu \mathcal{P}_1 \mathcal{P}_2-\mathcal{V}_\mu \partial^\mu\mathcal{P}_2 \mathcal{P}_1\right),  \notag\\
&\mathcal{L}_{\mathcal{V}_1 \mathcal{V}_2 \mathcal{P}}= & -g_{\scalebox{0.5}{$\mathcal{V}_1 \mathcal{V}_2 \mathcal{P}$}} \varepsilon^{\mu \nu \alpha \beta}\left(\partial_\mu \mathcal{V}_{1 \nu} \partial_\alpha \mathcal{V}_{2 \beta}\right) \mathcal{P},\notag\\
&\mathcal{L}_{\scalebox{0.5}{$\mathcal{T}_{cs(\bar{s})}\mathcal{P}_1\mathcal{P}_2$}}=&g_{\scalebox{0.5}{$\mathcal{T}_{cs(\bar{s})}\mathcal{P}_1\mathcal{P}_2$}} \mathcal{T}_{cs(\bar{s})}\mathcal{P}_1 \mathcal{P}_2+h.c.,\notag\\
&\mathcal{L}_{\scalebox{0.5}{$\mathcal{T}_{cs(\bar{s})}\mathcal{V}_1\mathcal{V}_2$}}=&g_{\scalebox{0.5}{$\mathcal{T}_{cs(\bar{s})}V_1V_2$}} \mathcal{T}_{cs(\bar{s})} \mathcal{V}_1^{\mu} \mathcal{V}_{2\mu}+h.c .
\end{eqnarray}
In the above expression, $\mathcal{P}$, $\mathcal{V}$, and $\mathcal{T}_{cs(\bar{s})}$ denote the light pseudoscalar meson, the vector meson, and singly charmed four-quark molecular state, respectively.
For the triangle diagrams $B^-\to \rho^- D^0 \xrightarrow{K^0} K^- T_{D^0 K^0}$, the short-distance process corresponds to tree level Wilson coefficients $a_1$, $a_2$ and CKM element $V_{cb}V_{ud}^*$, while the long-distance process corresponds to hadronic Lagrangian $\mathcal{L}_{\mathcal{V}\mathcal{P}_1\mathcal{P}_2}$ and $\mathcal{L}_{\mathcal{T}_{cs(\bar s)}\mathcal{P}_1\mathcal{P}_2}$, then the amplitude can be expressed as $\mathcal{M}_{1,2}^{DK}(\rho^-,D^0;K^0)_{V_{ud}}$.
Since hadrons are not point-like particles, we adopt the monopole-type form factor~\cite{Cheng:2004ru},
\begin{eqnarray}
\begin{aligned}
\mathcal{F}(k^2)=\left(\frac{m_{ex}^2-\Lambda^2}{k^2-\Lambda^2}\right)^2 .
\end{aligned}
\end{eqnarray}
$m_{ex}$ and $k$ represent the mass and momentum of the exchanged  particle respectively. The cutoff parameter is $\Lambda=m_{ex}+\alpha$. The parameter $\alpha=\eta\Lambda_{\mathrm{QCD}}$, with $1<\eta<4$, $\Lambda_{\mathrm{QCD}} = 0.218\,\mathrm{GeV}$~\cite{Han:2021azw}.

We write all the possible Cabibbo-allowed(CA) decay amplitudes, in which the amplitudes of PP configuration are
\begin{eqnarray}
\mathcal{M}(B^-\to K^-T^0_{DK}/T^0_{D_s\pi})&=&\mathcal{M}_{1,2}^{DK}(\rho^-,D^0;K^0)_{V_{ud}}+\mathcal{M}_{1,2}^{D_s\pi}(D^{*0},\pi^-;D_s^+)_{V_{ud}},\notag\\
\mathcal{M}(\overline B^0\to K^0T^0_{D\overline{K}}/T^0_{DK})&=&\mathcal{M}_2^{D\overline{K}}(\rho^0/\omega,D^0;\overline{K}^0)_{V_{ud}}+\mathcal{M}_1^{DK}(\rho^-,D^+;K^-)_{V_{ud}},\notag\\
\mathcal{M}(\overline B^0\to \overline{K}^0T^0_{DK}/T^0_{D_s\pi})&=&\mathcal{M}_2^{DK}(\rho^0/\omega,D^0;K^0)_{V_{ud}}+\mathcal{M}_1^{D_s\pi}(D^{*+},\pi^-;D_s^+)_{V_{ud}},\notag\\
\mathcal{M}(\overline B_s^0\to\pi^0 T^0_{DK}/T^0_{D_s\pi})&=&\mathcal{M}_2^{DK}(D^{*0},K^0;D^0)_{V_{ud}}+\mathcal{M}_2^{DK}(K^{*0},D^{0};K^0)_{V_{ud}}\notag\\
&&+\mathcal{M}_1^{D_s\pi}(D_s^{*+},\pi^-;D_s^+)_{V_{ud}}+\mathcal{M}_1^{D_s\pi}(\rho^-,D_s^{+};\pi^-)_{V_{ud}},\notag\\
\mathcal{M}(B^-\to D^-T^0_{D\overline{K}}/T^0_{DK})&=&\mathcal{M}_1^{D\overline{K}}(D_s^{*-},D^0;\overline{K}^0)_{V_{cs}}+\mathcal{M}_2^{DK}(J/\psi,D^+;K^-)_{V_{cs}},\notag\\
\mathcal{M}(B^+\to D^-T^{++}_{D_s\pi}/T^{++}_{DK})&=&\mathcal{M}_1^{D_s\pi}(\overline{D}^{*0},D_s^+;\pi^+)_{V_{cs}}+\mathcal{M}_2^{DK}(J/\psi,K^+;D^+)_{V_{cs}},\notag\\
\mathcal{M}(\overline B^0\to \overline{D}^0T^{0}_{DK}/T^{0}_{D\overline{K}})&=&\mathcal{M}_1^{DK}(D_s^{*-},D^+;K^-)_{V_{cs}}+\mathcal{M}_2^{D\overline{K}}(J/\psi,\overline{K}^0;D^0)_{V_{cs}},\notag\\
\mathcal{M}(B^0\to \overline{D}^0T^{0}_{D_s\pi}/T^{0}_{DK})&=&\mathcal{M}_1^{D_s\pi}(D^{*-},D_s^+;\pi^-)_{V_{cs}}+\mathcal{M}_2^{DK}(J/\psi,K^0;D^0)_{V_{cs}}, \notag\\
\end{eqnarray}
the amplitudes of VV configuration are given as,
\begin{eqnarray}
\mathcal{M}(B^-\to K^-T^0_{D^{*}K^{*}}/T^0_{D_s^{*}\rho})&=&\mathcal{M}_{1,2}^{D^{*}K^{*}}(\pi^-,D^{*0};K^{*0})_{V_{ud}}+\mathcal{M}_{1,2}^{D^{*}K^{*}}(\rho^-,D^{*0};K^{*0})_{V_{ud}}\notag\\
&&+\mathcal{M}_{1,2}^{D_s^{*}\rho}(D^{0},\rho^-;D_s^{*+})_{V_{ud}}+\mathcal{M}_{1,2}^{D_s^{*}\rho}(D^{*0},\rho^-;D_s^{*+})_{V_{ud}}, \notag\\
\mathcal{M}(\overline B^0\to K^0T^0_{D^{*}\overline{K}^{*}}/T^0_{D^{*}K^{*}})&=&\mathcal{M}_2^{D^{*}\overline{K}^{*}}(\pi^0/\eta,D^{*0};\overline{K}^{*0})_{V_{ud}}+\mathcal{M}_2^{D^{*}\overline{K}^{*}}(\rho^0/\omega,D^{*0};\overline{K}^{*0})_{V_{ud}}\notag\\
&&+\mathcal{M}_1^{D^{*}K^{*}}(\pi^-,D^{*+};K^{*-})_{V_{ud}}+\mathcal{M}_1^{D^{*}K^{*}}(\rho^-,D^{*+};K^{*-})_{V_{ud}},\notag\\
\mathcal{M}(\overline B^0\to \overline{K}^0T^0_{D^{*}K^{*}}/T^0_{D_s^{*}\rho})&=&\mathcal{M}_2^{D^{*}K^{*}}(\pi^0/\eta,D^{*0};K^{*0})_{V_{ud}}+\mathcal{M}_2^{D^{*}K^{*}}(\rho^0/\omega,D^{*0};K^{*0})_{V_{ud}}\notag\\
&&+\mathcal{M}_1^{D_s^{*}\rho}(D^{+},\rho^-;D_s^{*+})_{V_{ud}}+\mathcal{M}_1^{D_s^{*}\rho}(D^{*+},\rho^-;D_s^{*+})_{V_{ud}}, \notag\\
\mathcal{M}(\overline B_s^0\to\pi^0 T^0_{D^*K^*}/T^0_{D_s^*\rho})&=&\mathcal{M}_2^{D^*K^*}(D^{0},K^{*0};D^{*0})_{V_{ud}}+\mathcal{M}_2^{D^*K^*}(K^{0},D^{*0};K^{*0})_{V_{ud}}\notag\\
&&+\mathcal{M}_2^{D^*K^*}(D^{*0},K^{*0};D^{*0})_{V_{ud}}+\mathcal{M}_2^{D^*K^*}(K^{*0},D^{*0};K^{*0})_{V_{ud}}\notag\\
&&+\mathcal{M}_1^{D_s^*\rho}(D_s^{+},\rho^-;D_s^{*+})_{V_{ud}}+\mathcal{M}_1^{D_s^*\rho}(\pi^-,D_s^{*+};\rho^-)_{V_{ud}}\notag\\
&&+\mathcal{M}_1^{D_s^*\rho}(D_s^{*+},\rho^-;D_s^{*+})_{V_{ud}}+\mathcal{M}_1^{D_s^*\rho}(\rho^-,D_s^{*+};\rho^-)_{V_{ud}},\notag\\
\mathcal{M}(B^-\to D^-T^0_{D^{*}\overline{K}^{*}}/T^0_{D^{*}K^{*}})&=&\mathcal{M}_1^{D^{*}\overline{K}^{*}}(D_s^{-},D^{*0};\overline{K}^{*0})_{V_{cs}}+\mathcal{M}_1^{D^{*}\overline{K}^{*}}(D_s^{*-},D^{*0};\overline{K}^{*0})_{V_{cs}}\notag\\
&&+\mathcal{M}_2^{D^{*}K^{*}}(\eta_c,D^{*+};K^{*-})_{V_{cs}}+\mathcal{M}_2^{D^{*}K^{*}}(J/\psi,D^{*+};K^{*-})_{V_{cs}},\notag\\
\mathcal{M}(B^+\to D^-T^{++}_{D_s^{*}\rho}/T^{++}_{D^{*}K^{*}})&=&\mathcal{M}_1^{D_s^{*}\rho}(\overline{D}^{0},D_s^{*+};\rho^+)_{V_{cs}}+\mathcal{M}_1^{D_s^{*}\rho}(\overline{D}^{*0},D_s^{*+};\rho^+)_{V_{cs}}\notag\\
&&+\mathcal{M}_2^{D^{*}K^{*}}(\eta_c,K^{*+};D^{*+})_{V_{cs}}+\mathcal{M}_2^{D^{*}K^{*}}(J/\psi,K^{*+};D^{*+})_{V_{cs}},\notag\\
\mathcal{M}(\overline B^0\to \overline{D}^0T^{0}_{D^{*}K^{*}}/T^{0}_{D^{*}\overline{K}^{*}})&=&\mathcal{M}_1^{D^{*}K^{*}}(D_s^{-},D^{*+};K^{*-})_{V_{cs}}+\mathcal{M}_1^{D^{*}K^{*}}(D_s^{*-},D^{*+};K^{*-})_{V_{cs}}\notag\\
&&+\mathcal{M}_2^{D^{*}\overline{K}^{*}}(\eta_c,\overline{K}^{*0};D^{*0})_{V_{cs}}+\mathcal{M}_2^{D^{*}\overline{K}^{*}}(J/\psi,\overline{K}^{*0};D^{*0})_{V_{cs}},\notag\\
\mathcal{M}(B^0\to \overline{D}^0T^{0}_{D_s^{*}\rho}/T^{0}_{D^{*}K^{*}})&=&\mathcal{M}_1^{D_s^{*}\rho}(D^{-},D_s^{*+};\rho^{-})_{V_{cs}}+\mathcal{M}_1^{D_s^{*}\rho}(D^{*-},D_s^{*+};\rho^{-})_{V_{cs}}\notag\\
&&+\mathcal{M}_2^{D^{*}K^{*}}(\eta_c,K^{*0};D^{*0})_{V_{cs}}+\mathcal{M}_2^{D^{*}K^{*}}(J/\psi,K^{*0};D^{*0})_{V_{cs}},
\end{eqnarray}
Given the short-distance Wilson coefficients, the amplitude processes that depend solely on $a_2$ are suppressed. The Cabibbo-suppressed(CS) processes related to transitions $b\to c\bar u s$, $b\to c\bar c d$, and $b\to u \bar u d$ are given in Appendix.~\ref{sec:amplitudes}.

\section{Numerical Analysis: branching ratio and CP violation}
\label{sec:4}
\subsection{Branching ratios}
The input parameters, including decay constants, coupling coefficients~\cite{Shen:2019evi}, and Wilson coefficients, are listed in Tab.~\ref{tab:input_constants}. Furthermore, the B-meson transition form factors are evaluated using the z-series formula~\cite{McLean:2019qcx,Harrison:2021tol},
\begin{equation}
	f_i(q^2) = \frac{1}{1 - q^2/(m_{\text{pole}}^{(i)})^2}
	\sum_{k=0}^{N} a_k^{(i)} \, [z(q^2)]^k,
\end{equation}
where $f_i(q^2)$ can represent form factors of $B\to P$: $F_{1,2}(q^2)$ or $B\to V$: $V(q^2)$ and $A_{0,1,2}(q^2)$.
The variable $z(q^2)$ is defined as
\begin{equation}
	z(q^2) = \frac{\sqrt{t_+ - q^2} - \sqrt{t_+ - t_0}}
	{\sqrt{t_+ - q^2} + \sqrt{t_+ - t_0}},
\end{equation}
with $t_0 = q^2_{\max} = (m_B - m_{P/V})^2$ and
$t_+ = (m_B + m_{P/V})^2$. The pole masses are taken from~\cite{Detmold:2016pkz}. Then the fitted expansion parameters of form factors~\cite{Bharucha:2015bzk} are placed in Tab.~\ref{tab:formfactor}.

Finally, the production branching ratios from B meson decays for the PP and VV molecular state configurations are shown in Tab.~\ref{tab:result-PP} and Tab.~\ref{tab:result-VV}, respectively. In our calculations, we have taken into account the uncertainties associated with the strong coupling constant and the cutoff parameter $\alpha$, and adopted two different binding energies, $\epsilon=5$ MeV and $35$ MeV.  For the Cabibbo-allowed processes, we show the branching ratios of two PP or two VV molecular state configurations respectively. Among them,  $\overline B^0\to K^0 T_{D^{(*)0}\bar K^{(*)0}}^0$, $\overline B^0\to \overline K^0 T_{D^{(*)}K^{(*)}}^0$, $\overline B_s^0\to \pi^0 T_{D^{(*)}K^{(*)}}^0$, $B^-\to D^- T_{D^{(*)+}K^{(*)-}}^0$, $B^+\to D^- T_{D^{(*)+}K^{(*)+}}^{++}$, $\overline B^0\to \overline D^0 T_{D^{(*)}\bar K^{(*)}}^0$, and $B^0\to \overline D^0 T_{D^{(*)}K^{(*)}}^0$, proceed through the color-suppressed internal W-emission mechanism. Consequently, the weak decay amplitude involves only the Wilson coefficient $a_2$, generally leading to a relatively suppressed branching ratios. However, for process $\overline B^0\to \overline K^0 T_{DK}^0$, $B^+\to D^- T_{DK}^{++}$, and $\overline B_s^0\to \pi^0 T_{D^*K^*}^0$, the effect of final-state interactions (FSI) significantly enhances the branching fraction contributions. For $B^-$ $\to K^-T^0_{D_s^{+}\pi^-}/T^0_{D_s^{*+}\rho^-}$, although it receives contributions from both Wilson coefficients $a_1$ and $a_2$, the exchange of heavy D meson in the triangle diagram requires a large cutoff parameter, thereby suppressing the resulting branching ratio contribution. In the leading-order chiral effective Lagrangian, there is no direct coupling among three pseudoscalar mesons. Consequently, the triangle diagrams contributing to the production of PP molecular states are suppressed, leading to significantly smaller branching ratios compared to those of VV molecular states.

By comparing the results of the final-state interaction approach with those of the SU(3) symmetry analysis in Eq.(\ref{eq:CArelation}), we find that when the short-distance contributions are equal, the relations derived from the two methods are consistent with each other. Moreover, the ratios $R_1, R_2$ show a certain deviation between SU(3) flavor symmetry and final-state interaction results, indicating a non-negligible flavor symmetry breaking effect,
\begin{eqnarray}
&&R_1^a\Big(\frac{\Gamma(\overline B^0\to\overline{K}^0 T_{D^0K^0}^0)}{\Gamma(\overline B_s^0\to K^0 T_{D^0 \overline K^0}^0)}\Big)=18.07,\ R_1^{a'}\Big(\frac{\Gamma(\overline B^0\to\overline{K}^0 T_{D^{*0}K^{*0}}^0)}{\Gamma(\overline B_s^0\to K^0 T_{D^{*0}\overline K^{*0}}^0)}\Big)=7.65,\notag\\
&&R_1^b\Big(\frac{\Gamma(\overline B^0\to K^0 T_{D^0\overline K^0}^0)}{\Gamma(\overline B_s^0\to \overline{K}^0 T_{D^0K^0}^0)}\Big)=18.07,\ R_1^{b'}\Big(\frac{\Gamma(\overline B^0\to K^0 T_{D^{*0}\overline K^{*0}}^0)}{\Gamma(\overline B_s^0\to \overline{K}^0 T_{D^{*0}K^{*0}}^0)}\Big)=7.65,\notag\\
&&R_1^c\Big(\frac{\Gamma(B^- \to K^- T_{D_s^+\pi^-}^{0})}{\Gamma(B^- \to \pi^- T_{D^+K^-}^{0})}\Big)=7.20,\ R_1^{c'}\Big(\frac{\Gamma(B^- \to K^- T_{D_s^{*+}\rho^-}^{0})}{\Gamma(B^- \to \pi^- T_{D^{*+}K^{*-}}^{0})}\Big)=17.75,\notag\\
&&R_2^a\Big(\frac{\Gamma(B^-\to D_s^- T_{D_s^+\pi^-}^0)}{\Gamma(\overline B^0\to \overline{D}^0 T_{D^0 \overline K^0}^0)}\Big)=0.005,\ R_2^{a'}\Big(\frac{\Gamma(B^-\to D_s^- T_{D_s^{*+}\rho^{-}}^0)}{\Gamma(\overline B^0\to \overline{D}^0 T_{D^{*0} \overline K^{*0}}^0)}\Big)=0.173,\notag\\
&&R_2^b\Big(\frac{\Gamma(B_s^0\to \overline{D}^0 T_{D^0 \overline K^0}^0)}{\Gamma(B^+\to D^- T_{D^+K^+}^{++})}\Big)=0.005,\ R_2^{b'}\Big(\frac{\Gamma(B_s^0\to \overline{D}^0 T_{D^{*0}\overline K^{*0}}^0)}{\Gamma(B^+\to D^- T_{D^{*+}K^{*+}}^{++})}\Big)=0.065.
\end{eqnarray}
Significant deviations are observed between $R_1^{a',b',c}$, $R_2$ and the SU(3) analysis results $R_1^{SU(3)}=18.91, R_2^{SU(3)}=0.05$, suggesting pronounced flavor symmetry breaking in these processes. The approximation $R_2^{SU(3)}$ neglects the transition $b\to u\bar u d$ . Although $V_{ub}V_{ud}$ is smaller than $V_{cb}V_{cd}$, if the long-distance contributions of the transition $b\to u\bar u d$ are large, the ratio $R_2^{SU(3)}$ obtained via final-state interactions would exhibit a significant deviation. Our calculation shows that in these processes $B^-\to D_s^- T_{D_s^+\pi^-}^0$, $B^-\to D_s^- T_{D_s^{*+}\rho^-}^0$, $\overline B^0\to \overline D^0 T_{D^{(*)0}\bar K^{(*)0}}^0$, $B_s^0\to \overline D^0 T_{D^0\bar K^0}^0$, and $B^+\to D^-T_{D^+K^+}^{++}$ the contribution of $b\to u\bar u d$ cannot be neglected.

When considering the mixing of two molecular states (either both PP or both VV), we may define the physical representation $|X_{cs}(PP),\,I,\,I_{3}\rangle$ (or $|X_{cs}(VV),\,I,\,I_{3}\rangle$) with different isospin,
\begin{equation}
\begin{aligned}
&|X_{cs\bar{u}\bar{d}}^+,\,I=1,\, I_3=1\rangle=|D^+\overline{K}^0\rangle  ,\\
&|X_{cs\bar{u}\bar{d}}^0,\,I=1,\,I_3=0\rangle=\frac{1}{\sqrt{2}}\left(|D^+K^-\rangle-|D^0\overline{K}^0\rangle  \right)   ,\\
&|X_{cs\bar{u}\bar{d}}^-,\,I=1,\,I_3=-1\rangle=|D^0 K^-  \rangle  ,\\
&|X_{cs\bar{u}\bar{d}}^{'0},\,I=0,\,I_3=0\rangle=\frac{1}{\sqrt{2}}\left(|D^+K^-\rangle+|D^0\overline{K}^0\rangle  \right)   .\\
\end{aligned}
\end{equation}
The physical representation of the $|X_{c\overline{s}}(PP),\,I,\,I_{3}\rangle$ (or $|X_{c\overline{s}}(VV),\,I,\,I_{3}\rangle$) are defined as,
\begin{equation}
\begin{aligned}
&|X_{cu\bar{s}\bar{d}}^{++},\,I=1,\,I_3=1\rangle=\frac{1}{\sqrt{2}}\left(|D_s^+\pi^+\rangle+|D^+K^+\rangle \right)  ,\\
&|X_{cu\bar{s}\bar{u}}^{+},\,I=1,\,I_3=0\rangle=\frac{1}{\sqrt{2}}\left(|D_s^+\pi^0\rangle-|D^0K^+)\rangle \right)  ,\\
&|X_{cd\bar{s}\bar{d}}^{+},\,I=1,\,I_3=0\rangle=\frac{1}{\sqrt{2}}\left(|D_s^+\pi^0\rangle-|D^+K^0)\rangle \right)  ,\\
&|X_{cd\bar{s}\bar{u}}^{0},\,I=1,\,I_3=-1\rangle=\frac{1}{\sqrt{2}}\left(|D_s^+\pi^-\rangle+|D^0K^0\rangle \right)  ,\\
&|X_{cu\bar{s}\bar{u}}^{'+},\,I=0,\,I_3=0\rangle=\frac{1}{\sqrt{2}}\left(|D_s^+\pi^0\rangle+|D^0K^+)\rangle \right)  ,\\
&|X_{cd\bar{s}\bar{d}}^{'+},\,I=0,\,I_3=0\rangle=\frac{1}{\sqrt{2}}\left(|D_s^+\pi^0\rangle+|D^+K^0)\rangle \right)  .\\
\end{aligned}
\end{equation}
Here, $X_{cs(\bar s)}(PP)$ and  $X_{cs(\bar s)}(VV)$ represent the PP-mixed and VV-mixed states, respectively.
Then the production branching ratios of them in the binding energy $\epsilon=35$ MeV are deduced as,
\begin{equation}
\begin{alignedat}{3}
\mathcal{B}r\!\left(\overline B^{0}\to X_{cs\bar{u}\bar{d}}^0(PP)  \right)
&= 3.88\times10^{-7},\,\,
\mathcal{B}r\!\left(B^{+}\to X_{cu\bar{s}\bar{d}}^{++}(PP)\right)
&= 5.52\times10^{-8},\\
\mathcal{B}r\!\left(B^{-}\to X_{cs\bar{u}\bar{d}}^0(PP)  \right)
&= 5.58\times10^{-7},\,\,
\mathcal{B}r\!\left(B^{-}\to X_{cd\bar{s}\bar{u}}^0(PP)\right)
&= 1.45\times10^{-6},\\
\mathcal{B}r\!\left(\overline B^{0}\to X_{cs\bar{u}\bar{d}}^{'0}(PP)\right)
&= 3.18\times10^{-7}\,\,
\mathcal{B}r\!\left(B^{-}\to X_{cs\bar{u}\bar{d}}^{'0}(PP)\right)
&= 6.24\times10^{-7},\\
\mathcal{B}r\!\left(\overline B^0\to X_{cs\bar{u}\bar{d}}^0(VV) \right)
&= 7.16 \times 10^{-6},\,\,
\mathcal{B}r\!\left(B^+\to X_{cu\bar{s}\bar{d}}^{++}(VV)\right)
&= 7.67 \times 10^{-5},\\
\mathcal{B}r\!\left(B^-\to X_{cs\bar{u}\bar{d}}^0(VV) \right)
&= 4.39 \times 10^{-6},\,\,
\mathcal{B}r\!\left(B^-\to X_{cd\bar{s}\bar{u}}^0(VV)\right)
&= 1.33 \times 10^{-5},\\
\mathcal{B}r\!\left(\overline B^0\to X_{cs\bar{u}\bar{d}}^{'0}(VV)\right)
&= 1.08 \times 10^{-6},\,\,
\mathcal{B}r\!\left(B^-\to X_{cs\bar{u}\bar{d}}^{'0}(VV)\right)
&= 2.12 \times 10^{-5}.\\
\end{alignedat}
\end{equation}
The mass of $T_{cs(\bar{s})}$ can be approximately expressed in terms of the masses of its constituent mesons and the binding energy $\epsilon$. Meanwhile the experimentally observed $T_{cs0}^*(2870)^0$, $T_{c\bar s0}^*(2900)^0$, $T_{c\bar s0}^*(2900)^{++}$ states have masses much closer to the $D^*K^*/D^*\rho$ thresholds, while lying significantly above the $DK/D\pi$ thresholds, which suggesting that the states may all be molecular states with VV configurations. Furthermore, the branching ratios of processes $B^+ \to D^+ T_{\bar c\bar s0}(2870)^0$ have been measured experimentally,
\begin{eqnarray}
\mathcal{B}r(B^+\to D^+ T_{\bar c\bar s0}^*(2870)^0\to D^+ D^- K^+)=(1.2\pm0.5) \times 10^{-5}.
\end{eqnarray}
Assuming that the branching ratio of $T_{\bar c\bar s0}^*(2870)^0$ decay process is 50\%, then its production branching ratio should be around $\mathcal{B}r(B^+\to D^+ T_{\bar c\bar s0}(2870)^0)=2.4\times 10^{-5}$. This agrees with the branching ratio of the charge-conjugate process $\mathcal{B}r(B^-\to D^- T_{D^{*0}\overline K^{*0}})=2.5\times 10^{-5}$ obtained from our VV molecular calculation, implying that $T_{\bar c\bar s0}^*(2870)^0$ is a $\overline D^{*0}K^{*0}$ molecular state. Our calculations are essentially consistent with the results of ~\cite{Yu:2023avh,Chen:2020eyu}.
\begin{table}
\centering
\caption{Decay constants, Wilson coefficients, and coupling parameters used in this work. The unit of the decay constants is GeV. The Wilson coefficients are evaluated at \(\mu=\overline{m_b}(m_b)=4.40\,\mathrm{GeV}\) with \(\Lambda_{\overline{\mathrm{MS}}}^{(5)}=225\,\mathrm{MeV}\). The binding energies involved in the coupling constants containing $T_{cs(\overline{s})}$ are uniformly taken to be $20$ MeV.}\label{tab:input_constants}
\begin{tabular}{cccccccccc}
\hline\hline
~~~   $f_{D}$      ~~~ &$f_{D^*}$      ~~~&$f_{\pi}$           ~~~ &$f_{\rho}$ ~~~  &$f_{\eta_c}$ ~~~     & $f_{J\psi}$ ~~~  &$f_{D_s}$ ~~~     & $f_{D_s^*}$           \\
~~~   $0.206$     ~~~ &$0.252$      ~~~  &$0.13$       ~~~ &$0.205$        ~~~  &$0.387$    ~~~  & $0.418$     ~~~  &$0.247$    ~~~  & $0.272$   \\ \hline
~~~   $a_1$       ~~~ &$a_2$       ~~~  &$a_3$     ~~~ &$a_5$       ~~~  &$g_{\rho KK}$ ~~~  & $g_{D^* D_s K}$     ~~~  &$g_{K^*DD_s}$ ~~~  & $g_{\rho K^*K}$       \\
~~~   $1.07_{-0.11}^{+0.03}$     ~~~ &$0.28\pm0.02$      ~~~  &$-0.0045$       ~~~ &$-0.0032$        ~~~  &$4.27$    ~~~  & $4.27$                ~~~  &$4.27$    ~~~  & $9.08$             \\  \hline
~~~  $g_{D_s^*D^*K}$ ~~~  & $g_{D^*K^* D_s}$   &$g_{T_{c\bar{s}}D^0K^0}$ ~~~  & $g_{T_{c\bar{s}}D_s^+\pi^-}$  &$g_{T_{c\bar{s}}D^{*0}K^{*0}}$ ~~~  & $g_{T_{c\bar{s}}D_s^{*+}\rho^-}$   \\
~~~  $-7.07$    ~~~  & $-7.07$    & $2.46_{-1.61}^{+1.36}$   & $0.81_{-0.53}^{+0.43}$    & $7.34_{-2.64}^{+1.65}$    & $6.14_{-1.83}^{+0.97}$  \\    \hline  \hline
\end{tabular}
\end{table}
\begin{table}
\centering
\caption{Fitted expansion parameters of form factors for $B\to D^{(*)}$, $B\to \pi(\rho)$ and $B\to K^{(*)}$ transitions.}
\label{tab:formfactor}
\begin{tabular}{c c c c |c c  c |c c  c   }
\hline\hline
&$B\to D$ & $a_0$ & $a_1$
&$B\to \pi$  & $a_0$ & $a_1$
&$B\to K$  & $a_0$ & $a_1$ \\
\hline
& $F_1(0)$ & $0.666$ & $-0.260$
& $F_1(0)$ & $0.403$ & $-0.614$
& $F_1(0)$ & $0.332$ & $-0.926$ \\

& $F_2(0)$ & $0.666$ & $-3.236$
& $F_2(0)$ & $0.492$ & $-1.517$
& $F_2(0)$ & $0.332$ & $0.274$ \\
\hline\hline
&$B\to D^*$ & $a_0$ & $a_1$
&$B\to \rho$  & $a_0$ & $a_1$
&$B\to K^*$  & $a_0$ & $a_1$ \\
\hline
& $V(0)$ & $0.100$ & $-0.180$
& $V(0)$ & $0.33$ & $-0.86$
& $V(0)$ & $0.34$ & $-1.05$ \\

& $A_0(0)$ & $0.105$ & $-0.430$
& $A_0(0)$ & $0.360$ & $-0.830$
& $A_0(0)$ & $0.36$ & $-1.04$ \\

& $A_1(0)$ & $0.055$ & $-0.010$
& $A_1(0)$ & $0.26$ & $0.39$
& $A_1(0)$ & $0.27$ & $0.30$ \\

& $A_2(0)$ & $0.059$ & $-0.110$
& $A_2(0)$ & $0.23$ & $0.81$
& $A_2(0)$ & $0.23$ & $0.84$ \\
\hline\hline
\end{tabular}
\end{table}
\begin{table}\footnotesize
\centering
\caption{Branching ratios under the two possible PP molecular states. The uncertainties originate from variations in the strong coupling constants and the cutoff parameter $\alpha$.}
\label{tab:result-PP}
\renewcommand{\arraystretch}{1.3}
\begin{tabular}{c@{\hspace{3pt}}lccc@{\hspace{3pt}}lcccc}
\hline\hline
\multicolumn{2}{c}{\multirow{2}*{Process}}
&\multicolumn{2}{c}{Branching ratio($10^{-6}$)} & \multicolumn{2}{c}{\multirow{2}*{Process}}
&\multicolumn{2}{c}{Branching ratio($10^{-6}$)}\\
\cline{3-4} \cline{7-8}
\multicolumn{2}{c}{} & $\epsilon=5$ MeV & $\epsilon=35$ MeV &\multicolumn{2}{c}{} & $\epsilon=5$ MeV & $\epsilon=35$ MeV\\
\hline
$B^-$   & $\to K^-T^{0}_{D^0 K^0}$ & $0.96^{+0.48}_{-0.34}$ & $19.15^{+9.51}_{-6.90}$
& $B^{-}$ & $\to D^{-}T^0_{D^{0} \bar{K}^{0} }$ & $0.48^{+0.23}_{-0.17}$ & $9.37^{+4.52}_{-3.32}$  \\
& $\to K^-T^0_{D_s^+\pi^-}$ & $0.0003^{+0.0001}_{-0.0001}$ & $0.0036^{+0.0019}_{-0.0013}$ &
& $\to D^{-}T^0_{D^+ K^- }$  & $0.0004^{+0.0002}_{-0.0002}$ & $0.0063^{+0.0037}_{-0.0025}$    \\
$\overline B^{0}$ & $\to K^{0}T^0_{D^{0}\overline{K}^{0}}$ & $0.009^{+0.005}_{-0.003}$ & $0.171^{+0.085}_{-0.062}$
& $B^{+}$ & $\to D^{-}T^{++}_{D_s^{+}\pi^{+}}$ & $0.002^{+0.001}_{-0.001}$ & $0.035^{+0.018}_{-0.013}$\\
 & $\to K^{0}T^0_{D^{+}K^{-}}$ & $0.92^{+0.45}_{-0.33}$ & $17.69^{+8.80}_{-6.37}$ &
& $\to D^{-}T^{++}_{D^{+} K^{+}}$ & $0.004^{+0.002}_{-0.002}$ & $0.080^{+0.044}_{-0.031}$\\
$\overline B_s^0$  & $\to \pi^0 T^{0}_{D^0K^0}$  & $0.005^{+0.002}_{-0.002}$ & $0.078^{+0.036}_{-0.026}$
& $\overline B^0$ & $\to \overline D^0 T^0_{D^0 \overline K^0}$ & $0.0004^{+0.0002}_{-0.0001}$ & $0.0060^{+0.0035}_{-0.0024}$\\
& $\to \pi^0 T^0_{D_s^+ \pi^-}$ & $0.05^{+0.02}_{-0.02}$ & $0.57^{+0.39}_{-0.24}$ &
& $\to \overline D^0 T^0_{D^+ K^-}$ & $0.43^{+0.21}_{-0.15}$ & $8.47^{+4.10}_{-3.01}$\\
$\overline B^0$ & $\to \overline K^0 T^0_{D^0 K^0}$ & $0.009^{+0.005}_{-0.003}$ & $0.17^{+0.09}_{-0.06}$
& $B^0$ & $\to \overline D^0 T^0_{D^0 K^0}$ & $0.0004^{+0.0002}_{-0.0001}$ & $0.0060^{+0.0035}_{-0.0024}$ \\
& $\to \overline K^0 T^0_{D_s^+ \pi^-}$ & $2.16^{+1.44}_{-0.92}\times10^{-5}$ & $0.0004^{+0.0003}_{-0.0002}$ &    & $\to \overline D^0 T^0_{D_s^+ \pi^-}$ & $0.002^{+0.001}_{-0.001}$ & $0.031^{+0.016}_{-0.011}$ \\ \hline
$B^-$ & $\xrightarrow{CS} \pi^- T^0_{D^0 \overline K^0}$ & $0.04^{+0.02}_{-0.01}$ & $0.63^{+0.30}_{-0.22}$
& $B^+$   & $\xrightarrow{CS} D_s^- T^{++}_{D_s^{+} \pi^+}$ & $3.77^{+2.37}_{-1.57}\times10^{-6}$ & $6.84^{+4.37}_{-2.88}\times10^{-5}$ &&\\
& $\xrightarrow{CS} \pi^- T^0_{D^+ K^-}$ & $4.04^{+1.79}_{-1.33}\times10^{-5}$ & $0.0005^{+0.0002}_{-0.0002}$ &
& $\xrightarrow{CS} D_s^- T^{++}_{D^{+} K^+}$ & $0.002^{+0.001}_{-0.001}$ & $0.033^{+0.015}_{-0.011}$ \\
$\overline B^0$ & $\to \pi^0 T^0_{D^0 \overline{K}^0}$ & $0.0003^{+0.0001}_{-0.0001}$ & $0.0046^{+0.0021}_{-0.0016}$
&$\overline B_s^0$ & $\xrightarrow{CS} \overline D^0 T^0_{D^0 K^0}$   & $1.97^{+1.43}_{-0.90}\times10^{-6}$ & $5.73^{+3.60}_{-2.39}\times10^{-5}$   \\
& $\to \pi^0 T^0_{D^+ K^-}$ & $0.02^{+0.01}_{-0.01}$ & $0.29^{+0.14}_{-0.10}$ &
& $\xrightarrow{CS} \overline D^0 T^{0}_{D_s^{+} \pi^-}$ & $0.0011^{+0.0009}_{-0.0005}$ & $0.018^{+0.016}_{-0.010}$ \\
$\overline B_s^0$ & $\xrightarrow{CS} K^0 T^0_{D^0 \overline K^0}$ & $0.0006^{+0.0003}_{-0.0002}$ & $0.0095^{+0.0044}_{-0.0032}$
& $B_s^0$ & $\xrightarrow{CS} \overline{D}^0 T^0_{D^0 \overline {K}^0}$ & $2.46^{+1.50}_{-1.01}\times10^{-5}$ & $0.0004^{+0.0003}_{-0.0002}$\\
& $\xrightarrow{CS} K^0 T^0_{D^+ K^-}$ & $6.29^{+3.43}_{-2.35}\times10^{-6}$ & $9.20^{+5.44}_{-3.64}\times10^{-5}$ &
& $\xrightarrow{CS} \overline D^{0} T^{0}_{D^+ K^-}$ & $0.0007^{+0.0004}_{-0.0003}$ & $0.014^{+0.008}_{-0.005}$ \\
$\overline B_s^0$ & $\xrightarrow{CS} \overline K^0 T^0_{D^0 K^0}$ & $0.0006^{+0.0003}_{-0.0002}$ & $0.0095^{+0.0044}_{-0.0032}$
& $B^-$ & $\xrightarrow{CS} D_s^- T^0_{D_s^+ \pi^-}$ & $1.85^{+1.16}_{-0.77}\times10^{-6}$ & $3.21^{+2.06}_{-1.35}\times10^{-5}$ \\
& $\xrightarrow{CS} \overline K^0 T^0_{D_s^+ \pi^-}$ & $0.003^{+0.002}_{-0.001}$ & $0.045^{+0.040}_{-0.023}$ &
& $\xrightarrow{CS} D_s^- T^{0}_{D^{0} K^0}$ & $0.04^{+0.02}_{-0.01}$ & $0.75^{+0.33}_{-0.25}$ \\
\hline
\end{tabular}
\end{table}
\begin{table}
\centering
\caption{Branching ratios under the two possible VV molecular states. The uncertainties originate from variations in the strong coupling constants and the cutoff parameter $\alpha$.}
\label{tab:result-VV}
\renewcommand{\arraystretch}{1.3}
\setlength{\tabcolsep}{5pt}
\begin{tabular}{c@{\hspace{3pt}}lccc@{\hspace{3pt}}lcccc}
\hline\hline
\multicolumn{2}{c}{\multirow{2}*{Process}}
&\multicolumn{2}{c}{Branching ratio($10^{-4}$)} & \multicolumn{2}{c}{\multirow{2}*{Process}}
&\multicolumn{2}{c}{Branching ratio($10^{-4}$)}\\
\cline{3-4} \cline{7-8}
\multicolumn{2}{c}{} & $\epsilon=5$ MeV & $\epsilon=35$ MeV &\multicolumn{2}{c}{} & $\epsilon=5$ MeV & $\epsilon=35$ MeV\\
\hline
$B^-$   & $\to K^-T^{0}_{D^{*0} K^{*0}}$ & $0.79^{+0.35}_{-0.26}$ & $2.75^{+1.24}_{-0.92}$
&$B^{-}$ & $\to D^{-}T^0_{D^{*0}\overline {K}^{*0}}$ & $0.07^{+0.03}_{-0.02}$ & $0.25^{+0.11}_{-0.08}$  \\
& $\to K^-T^0_{D_s^{*+}\rho^-}$  & $0.35^{+0.26}_{-0.16}$ & $0.96^{+0.70}_{-0.43}$ &
& $\to D^{-}T^0_{D^{*+} K^{*-}}$      & $0.01^{+0.01}_{-0.01}$ & $0.04^{+0.04}_{-0.02}$\\
$\overline B^{0}$ & $\to K^{0}T^0_{D^{*0}\overline {K}^{*0}}$ & $0.03^{+0.01}_{-0.01}$ & $0.08^{+0.04}_{-0.03}$
& $B^{+}$ & $\to D^{-}T^{++}_{D_s^{*+}\rho^{+}}$ & $0.65^{+0.42}_{-0.27}$ & $2.57^{+1.63}_{-1.08}$  \\
& $\to K^{0}T^0_{D^{*+}K^{*-}}$ & $0.02^{+0.01}_{-0.01}$ & $0.07^{+0.03}_{-0.02}$ &
& $\to D^- T^{++}_{D^{*+} K^{*+}}$     &  $0.01^{+0.01}_{-0.01}$ & $0.04^{+0.04}_{-0.02}$ \\
$\overline B_s^0$ & $\to \pi^0 T^0_{D^{*0}K^{*0}}$ & $0.25^{+0.12}_{-0.08}$ & $0.86^{+0.41}_{-0.30}$
& $\overline B^0$ & $\to \overline D^0 T^{0}_{D^{*0} \overline K^{*0}}$ & $0.01^{+0.01}_{-0.01}$ & $0.03^{+0.04}_{-0.02}$  \\
& $\to \pi^0 T^0_{D_s^{*+} \rho^-}$ & $0.007^{+0.003}_{-0.002}$ & $0.028^{+0.011}_{-0.085}$ &
& $\to \overline D^0 T^{0}_{D^{*+} K^{*-}}$ & $0.06^{+0.03}_{-0.02}$ & $0.23^{+0.11}_{-0.08}$ \\
$\overline B^0$ & $\to \overline K^0 T^0_{D^{*0} K^{*0}}$ & $0.03^{+0.01}_{-0.01}$ & $0.08^{+0.04}_{-0.03}$
& $B^0$ & $\to \overline D^0 T^0_{D^{*0} K^{*0}}$ & $0.01^{+0.01}_{-0.01}$ & $0.03^{+0.04}_{-0.02}$ \\
& $\to \overline K^0 T^0_{D_s^{*+} \rho^{*-}}$ & $0.26^{+0.15}_{-0.10}$ & $0.69^{+0.40}_{-0.28}$  &
& $\to \overline D^0 T^0_{D_s^{*+} \rho^{*-}}$ & $0.60^{+0.39}_{-0.25}$ & $2.38^{+1.51}_{-1.00}$\\ \hline
$B^-$ & $\xrightarrow{CS} \pi^- T^0_{D^{*0} \overline K^{*0}}$ & $0.007^{+0.003}_{-0.002}$ & $0.020^{+0.009}_{-0.006}$
& $B^+$   & $\xrightarrow{CS} D_s^- T^{++}_{D_s^{*+} \rho^+}$  & $0.002^{+0.003}_{-0.001}$ & $0.006^{+0.007}_{-0.003}$ \\
& $\xrightarrow{CS} \pi^- T^0_{D^{*+} K^{*-}}$ & $0.02^{+0.01}_{-0.01}$ & $0.05^{+0.03}_{-0.02}$ &
& $\xrightarrow{CS} D_s^- T^{++}_{D^{*+} K^{*+}}$ & $0.12^{+0.06}_{-0.04}$ & $0.47^{+0.21}_{-0.16}$ \\
$\overline B^0$   & $\xrightarrow{CS} \pi^0 T^{0}_{D^{*0} \overline K^{*0}}$ & $0.005^{+0.002}_{-0.002}$ & $0.016^{+0.008}_{-0.006}$
& $\overline B_s^0$ & $\xrightarrow{CS} \overline D^0 T^{0}_{D^{*0} K^{*0}}$ & $0.0008^{+0.0009}_{-0.0004}$ & $0.0022^{+0.0024}_{-0.0012}$ \\
& $\xrightarrow{CS} \pi^0 T^{0}_{D^{*+} K^{*-}}$ & $0.008^{+0.004}_{-0.003}$ & $0.022^{+0.012}_{-0.009}$    &
& $\xrightarrow{CS} \overline D^0 T^{0}_{D_s^{*+} \rho^-}$ & $0.0008^{+0.0001}_{-0.0001}$ & $0.0036^{+0.0008}_{-0.0008}$ \\
$\overline B_s^0$ & $\xrightarrow{CS} K^0 T^0_{D^{*0} \overline K^{*0}}$ & $0.003^{+0.001}_{-0.001}$ & $0.011^{+0.005}_{-0.004}$
&  $B_s^0$ & $\xrightarrow{CS} \overline{D}^{0} T^0_{D^{*0} \overline {K}^{*0}}$ & $0.0008^{+0.0008}_{-0.0004}$ & $0.0022^{+0.0024}_{-0.0012}$   \\
& $\xrightarrow{CS} K^0 T^0_{D^{*+} K^{*-}}$ & $0.014^{+0.009}_{-0.006}$ & $0.041^{+0.027}_{-0.017}$ &
& $\xrightarrow{CS} \overline D^{0} T^{0}_{D^{*+} K^{*-}}$ & $0.20^{+0.09}_{-0.07}$ & $0.76^{+0.35}_{-0.26}$ \\
$\overline B_s^0$ & $\xrightarrow{CS} \overline K^0 T^0_{D^{*0} K^{*0}}$ & $0.003^{+0.001}_{-0.001}$ & $0.011^{+0.005}_{-0.004}$
& $B^-$ & $\xrightarrow{CS} D_s^- T^0_{D_s^{*+} \rho^-}$ & $0.002^{+0.003}_{-0.001}$ & $0.006^{+0.007}_{-0.003}$ \\
& $\xrightarrow{CS} \overline K^0 T^0_{D_s^{*+} \rho^-}$ & $0.002^{+0.001}_{-0.001}$ & $0.007^{+0.003}_{-0.002}$ &
& $\xrightarrow{CS} D_s^- T^{0}_{D^{*0} K^{*0}}$ & $0.003^{+0.001}_{-0.001}$ & $0.010^{+0.004}_{-0.003}$ \\
\hline
\end{tabular}
\end{table}
\subsection{CP violation}
\label{sec:5}
Study of rare decays of B meson provides an important avenue for probing the underlying mechanisms of CP violation. In this work, we aim to investigate the possible CP-violating effects arising in the production processes. Owing to the interference between tree and penguin diagrams or different penguin diagrams, as well as the strong phase introduced by final-state interactions, direct CP violation may be sizable. Direct CP violation is defined as,
\begin{equation}
\begin{aligned}
A_{CP}^{dir}=\frac{\Gamma(B\to f)-\Gamma(\overline{B}\to \overline{f})}{\Gamma(B\to f)+\Gamma(\overline{B}\to \overline{f})}.
\end{aligned}
\end{equation}
For the $b\to c\bar c s$ transition, the contribution of tree diagram dominates while the penguin contribution is suppressed, leading to a suppressed CP violation. In contrast, for the $b\to c\bar cd$ and $b\to u\bar u d$ transitions, the penguin diagrams provide a large weak phase from $V_{td}$ and $V_{ub}$. When combined with the strong phase arising from final-state interactions, the CP violation can be significantly enhanced. Thus we restrict our numerical calculations to Cabibbo-suppressed channels $b\to c\bar{c}d$ and $b\to u\bar u d$ in the work.

Although we find that $T_{cs/\bar s}$s prefer the VV molecular states, we still analyze the CP violation effects for the two molecular states separately.
The branching ratios of the PP molecular states are smaller than that of the VV molecular states, but the CP violation behavior is different. For the PP molecular-state, the direct CP asymmetries of those decay channels are sizable, we present here processes with appreciable CP violation.
\begin{eqnarray}
A_{CP}(B^-\to D_s^- T^0_{D_s^{+}\pi^-})=&0.087^{+0.003}_{-0.003},\notag\\
A_{CP}(B_s^0\to \overline{D}^{0} T^0_{D^{0}\overline{K}^{0}})=&0.066^{+0.004}_{-0.005},\notag\\
A_{CP}(B^+\to D_s^- T^{++}_{D_s^{+}\pi^+})=&-0.087^{+0.003}_{-0.003},\notag\\
A_{CP}(\overline B_s^0\to \overline D^0 T^{0}_{D^{0}K^{0}})=&-0.066^{+0.005}_{-0.004}.
\end{eqnarray}
Under the VV molecular-state configuration, the relevant direct CP violation is around $\mathcal{O}(10^{-3})$,
\begin{eqnarray}
A_{CP}(B^-\to D_s^- T^0_{D_s^{*+}\rho^-})=&0.0065^{+0.0001}_{-0.0002},\notag\\
A_{CP}(B_s^0\to \overline{D}^{0} T^0_{D^{*0}\overline{K}^{*0}})=&-0.0076^{+0.0019}_{-0.0011},\notag\\
A_{CP}(B^+\to D_s^- T^{++}_{D_s^{*+}\rho^+})=&-0.0065^{+0.0002}_{-0.0001},\notag\\
A_{CP}(\overline B_s^0\to \overline D^0 T^{0}_{D^{*0}K^{*0}})=&0.0076^{+0.0011}_{-0.0019}.
\end{eqnarray}
\begin{figure}[htbp]
\centering
  \begin{minipage}[t]{0.49\linewidth}
    \centering
    \includegraphics[width=0.95\textwidth]{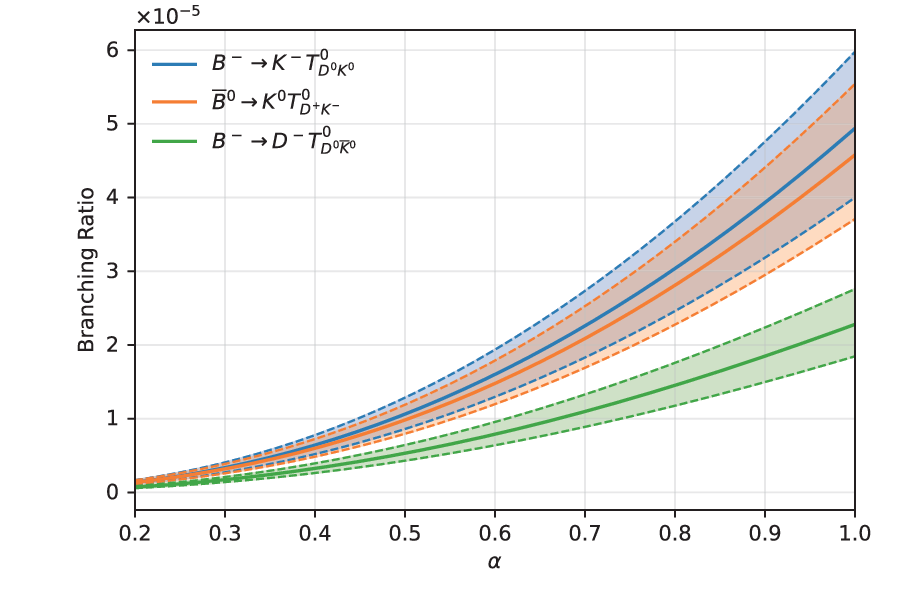}
\end{minipage}
  \begin{minipage}[t]{0.49\linewidth}
    \centering
    \includegraphics[width=0.95\textwidth]{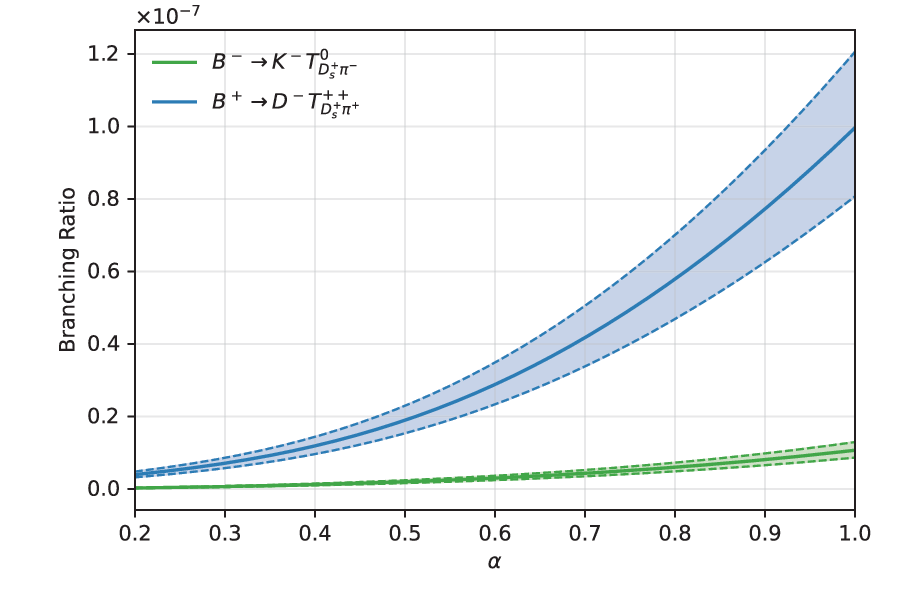}
\end{minipage}
\caption{The branching ratios of the individual production channels within the PP molecular configuration as a function of the model parameter $\alpha$, where the parameter varies from 0.2 to 1.0 GeV. The binding energy $\epsilon$ is taken to be 35 MeV. The solid lines correspond to the central values, while the shaded bands denote the theoretical uncertainties.}
\label{Fig:4}
\end{figure}
\begin{figure}[htbp]
\centering
  \begin{minipage}[t]{0.49\linewidth}
    \centering
    \includegraphics[width=0.95\textwidth]{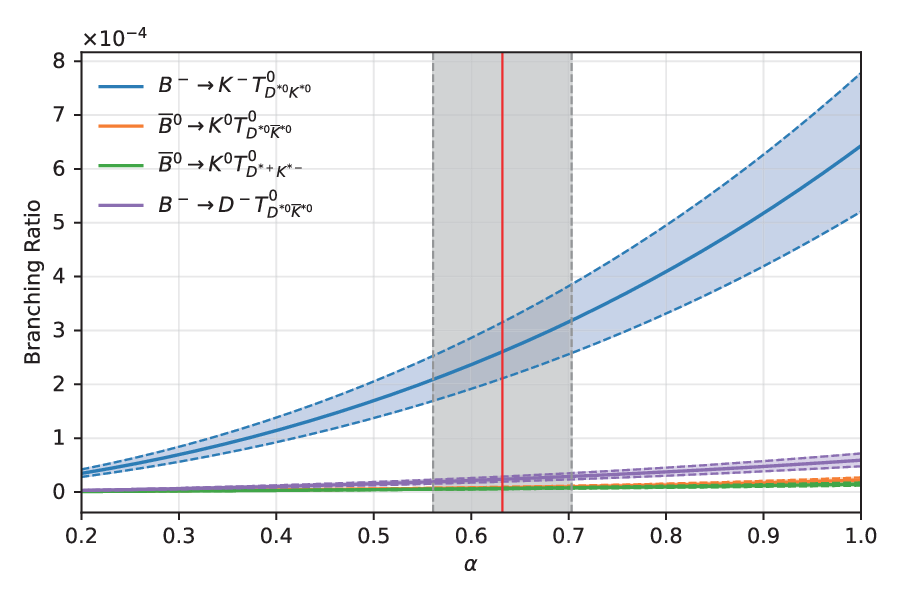}
\end{minipage}
  \begin{minipage}[t]{0.49\linewidth}
    \centering
    \includegraphics[width=0.95\textwidth]{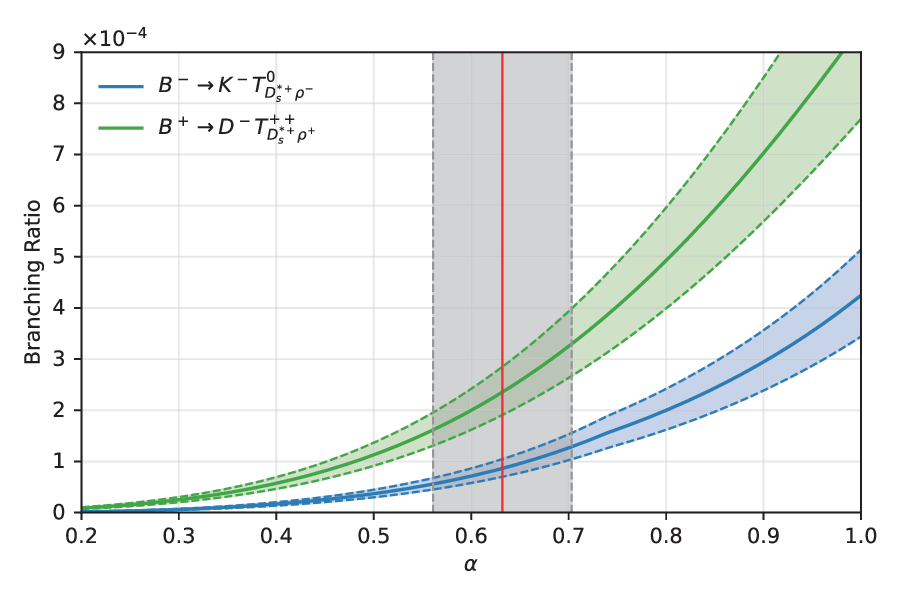}
\end{minipage}
\caption{The branching ratios of the individual production channels within the VV molecular configuration as a function of the model parameter $\alpha$, where the parameter varies from 0.2 to 1.0 GeV. The binding energy $\epsilon$ is taken to be 35 MeV. The solid lines correspond to the central values, while the shaded bands denote the theoretical uncertainties. The shadow indicates that the branching ratio of $B^- \to D^- T^0_{D^{*0}\overline{K}^{*0}}$ within this range is consistent with the experimental results within uncertainties. The red line denotes the central value of the shadow region.}
\label{Fig:5}
\end{figure}
\section{Conclusions}
\label{sec:6}
In this work, we investigate the production of scalar $D^{(*)}K^{(*)}$ and $D^{(*)}\pi(\rho)$ molecular states from B meson decays within the SU(3) flavor symmetry analysis and final-state interaction approach. We calculate the production branching ratios of these states. The results show that processes with the VV molecular configuration generally yield larger branching ratios than those with the corresponding PP molecular configuration, supporting the interpretation that the recently observed $T_{\bar c\bar s0}^*(2870)^0$ possesses a VV molecular structure. In addition, we evaluate the possible CP violation in the production processes. Owing to the interference between different diagrams, together with the enhancement from final-state interactions, a sizable amount of CP violation is obtained. The results of our analysis are expected to offer valuable insights for both future experimental endeavors and theoretical investigations.

\section*{Acknowledgements}
\appendix
\section{}\label{sec:coupling constant}
\label{app:1}
\begin{figure}[ht]
\centering
\includegraphics[scale=0.5]{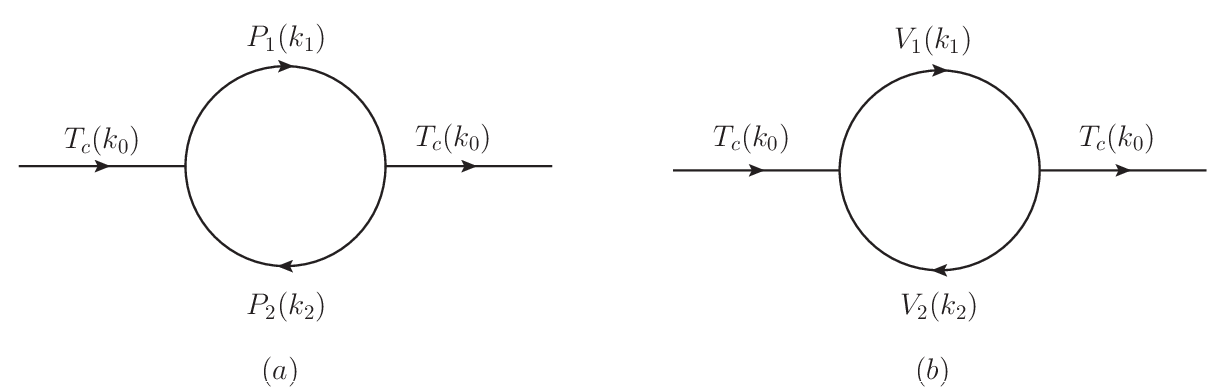}
\caption{The self-energy diagrams of $T_{cs(\bar{s})}$ which are regarded as a molecular state of either two pseudoscalar mesons (diagram (a)) or two vector mesons (diagram (b)).}
\label{Fig:3}
\end{figure}

To find out the couplings between $T_{cs(\bar s)}$ and its meson-meson constituents, we use the effective Lagrangian which including the correlation function $\phi(y^2)$ to characterize the interaction. The Lagrangian describe the interaction between $T_{cs(\bar s)}$ with two pseudoscalar mesons and two vector mesons are respectively given as~\cite{Faessler:2007us}:
\begin{eqnarray}
\label{TPP}
&&\mathcal{L}_{\scalebox{0.5}{$\mathcal{T}_{cs(\bar{s})}\mathcal{P}_1\mathcal{P}_2$}}(x)=g_{\scalebox{0.5}{$\mathcal{T}_{cs(\bar{s})}\mathcal{P}_1\mathcal{P}_2$}} \mathcal{T}_{cs(\bar{s})}(x) \int d y \Phi\left(y^2\right) \mathcal{P}_1\left(x+\omega_1 y\right) \mathcal{P}_2\left(x-\omega_2 y\right)+h.c.,\\
&&\mathcal{L}_{\scalebox{0.5}{$\mathcal{T}_{cs(\bar{s})}\mathcal{V}_1\mathcal{V}_2$}}(x)=g_{\scalebox{0.5}{$\mathcal{T}_{cs(\bar{s})}V_1V_2$}} \mathcal{T}_{cs(\bar{s})}(x) \int d y \Phi\left(y^2\right) \mathcal{V}_1^{\mu}\left(x+\omega_{1} y\right) \mathcal{V}_{2\mu}\left(x-\omega_{2} y\right)+h.c. .
\end{eqnarray}
Where $\omega_{i}=\frac{m_i}{m_i+m_j}$ ($i=1, 2$) is a kinematic parameter, with $m_{i}$ and $m_j$ denoting the masses of $\mathcal{P}_1(\mathcal{V}_1)$, $\mathcal{P}_2(\mathcal{V}_2)$ respectively. The correlation function $\Phi\left(y^2\right)$ describes the distribution of the two constituent mesons and ensures that the Feynman diagrams remain ultraviolet finite, which adopting a Gaussian form~\cite{Branz:2008cb}:
\begin{eqnarray}
\begin{aligned}
\label{eq:20}
\Phi\left(y^2\right)=\int \frac{d^4 p}{(2 \pi)^4} e^{-i p y} \widetilde{\Phi}\left(-p^2\right) ,\, \widetilde{\Phi}\left(p^2\right) \doteq \exp \left(-p^2 / \Lambda_{T_{cs(\bar{s})}}^2\right).
\end{aligned}
\end{eqnarray}
$\Lambda_{T_{cs(\bar{s})}}$ is the size parameter characterizing the distribution of the components inside the molecular state, which is set to 1.0 GeV. Using the Weinberg compositeness condition~\cite{Faessler:2007us,Huang:2018wgr},
\begin{eqnarray}
\begin{aligned}
\label{Eq14}
Z_{T_{cs(\bar{s})}}=1-\left.\frac{d \Sigma_{T_{cs(\bar{s})}}}{dp^2}\right|_{p^2=m_{T_{cs(\bar{s})}}^2}=0.
\end{aligned}
\end{eqnarray}
The explicit form of the self-energy diagram Fig.~\ref{Fig:3} are given as:
\begin{eqnarray}
\begin{aligned}
\Sigma_{T_{cs(\bar{s})}}(k_0)& = g_{\scalebox{0.5}{$T_{cs(\bar{s})}P_1P_2$}}^2 \int \frac{d^4 k_1}{(2 \pi)^4\mathcal{i}} \Phi^2\left(-\left(k_1-k_0 \omega_{1}\right)^2\right)\frac{1}{k_1^2-m_{P_1}^2} \frac{1}{\left(k_0-k_1\right)^2-m_{P_2}^2},  \\
\Sigma_{T_{cs(\bar{s})}}(k_0)& = g_{\scalebox{0.5}{$T_{cs(\bar{s})}V_1V_2$}}^2 \int \frac{d^4 k_1}{(2 \pi)^4\mathcal{i}} \Phi^2\left(-\left(k_1-k_0 \omega_{1}\right)^2\right) \\
& \times \frac{-g^{\mu \nu}+k_1^\mu k_{1}^{\nu} / m_{V_1}^2}{k_1^2-m_{V_1}^2} \frac{-g_{\mu \nu}+\left(k_0-k_1\right)_\mu\left(k_0-k_1\right)_\nu / m_{V_2}^2}{\left(k_0-k_1\right)^2-m_{V_2}^2},
\end{aligned}
\end{eqnarray}
where $k_0$, $k_1$, and $k_2$  represent the four-momenta of the corresponding hadrons. Then the explicit form of the coupling constant $g_{T_{cs(\bar{s})}P_1P_2}$ and $g_{T_{cs(\bar{s})}V_1V_2}$ are
\begin{eqnarray}
\begin{aligned}
\frac{1}{g_{\scalebox{0.5}{$T_{cs(\bar{s})}PP$}}^2}=&\int^1_0d\alpha \int^\infty_0d\beta \frac{-1}{8\pi^2}\frac{1}{(-1-\beta)^2}\frac{\beta}{\Lambda^2}e^{\frac{-2\Delta}{\Lambda^2}}\left(-\frac{m_2^2}{(m_1+m_2)^2}-(1-\alpha)\beta-\frac{1}{4}\frac{\left(2(1-\alpha)\beta-\frac{-2m_2}{m_1+m_2}\right)^2}{(-1-\beta)}  \right)  ,   \\
\frac{1}{g_{\scalebox{0.5}{$T_{cs(\bar{s})}VV$}}^2}=&\frac{\partial }{\partial p^2} \int^1_0d\alpha \int^\infty_0d\beta \frac{\beta}{16\pi^2}\left(4\frac{e^{\frac{-2\Delta}{\Lambda^2}}}{(-1-\beta)^2}-\frac{1}{m^2}\frac{\Lambda^2}{2}A_1-\frac{1}{m_1^2}\left(p^2\frac{e^{\frac{-2\Delta}{\Lambda^2}}}{(-1-\beta)^2}-2\frac{\Lambda^2}{2}A_2+\frac{\Lambda^2}{2}A_1  \right)  \right.  \\
&  \left.  +\frac{1}{m_1^2m_2^2}\left(\frac{\Lambda^4}{4}A_4-2\frac{\Lambda^4}{4}A_5+\frac{\Lambda^4}{4}A_3  \right) \right) ,  \\
\Delta=&-\frac{m_2^2}{(m_1+m_2)^2}p^2+\alpha\beta m_2^2-(1-\alpha)\beta(p^2-m_1^2)-\frac{1}{4}\frac{\left(2(1-\alpha)\beta-\frac{-2m_2}{m_1+m_2}\right)^2}{(-1-\beta)}p^2 ,  \\
A_1=&2\frac{e^{\frac{-2\Delta}{\Lambda^2}}}{(-1-\beta)^3}-\frac{1}{4}\frac{-2}{\Lambda^2}\frac{e^{\frac{-2\Delta}{\Lambda^2}}}{(-1-\beta)^4}\left(2(1-\alpha)\beta-\frac{-2m_2}{m_1+m_2}\right)^2p^2  ,  \\
A_2=&\frac{2}{4} \frac{e^{\frac{-2\Delta}{\Lambda^2}}}{(-1-\beta)^3}\left(2(1-\alpha)\beta-\frac{-2m_2}{m_1+m_2}\right) p^2 \frac{-2}{\Lambda^2} ,  \\
A_3=&6\frac{e^{\frac{-2\Delta}{\Lambda^2}}}{(-1-\beta)^4}-\frac{2}{4}\frac{-2}{\Lambda^2}\frac{e^{\frac{-2\Delta}{\Lambda^2}}}{(-1-\beta)^5}\left(2(1-\alpha)\beta-\frac{-2m_2}{m_1+m_2}\right)^2p^2-\frac{4}{4}\frac{e^{\frac{-2\Delta}{\Lambda^2}}}{(-1-\beta)^5}  \\
&\left(2(1-\alpha)\beta-\frac{-2m_2}{m_1+m_2}\right)^2p^2\frac{-2}{\Lambda^2}
+\frac{1}{16}\left(2(1-\alpha)\beta-\frac{-2m_2}{m_1+m_2}\right)^4p^4\frac{e^{\frac{-2\Delta}{\Lambda^2}}}{(-1-\beta)^6}\left(\frac{-2}{\Lambda^2} \right)^2 ,\\
A_4=&\frac{1}{4}\frac{-2e^{\frac{-2\Delta}{\Lambda^2}}}{(-1-\beta)^3}p^2\frac{-2}{\Lambda^2} +\frac{4}{16}\left(2(1-\alpha)\beta-\frac{-2m_2}{m_1+m_2}\right)^2p^4\frac{e^{\frac{-2\Delta}{\Lambda^2}}}{(-1-\beta)^4}\left(\frac{-2}{\Lambda^2} \right)^2  , \\
A_5=&\frac{2}{4}\frac{2e^{\frac{-2\Delta}{\Lambda^2}}}{(-1-\beta)^4}\left(2(1-\alpha)\beta-\frac{-2m_2}{m_1+m_2}\right)  p^2\frac{-2}{\Lambda^2}+\frac{2}{4}\left(2(1-\alpha)\beta-\frac{-2m_2}{m_1+m_2}\right)\frac{e^{\frac{-2\Delta}{\Lambda^2}}}{(-1-\beta)^4}p^2\frac{-2}{\Lambda^2}   \\
&-\frac{2}{16}\left(2(1-\alpha)\beta-\frac{-2m_2}{m_1+m_2}\right)^3 \frac{e^{\frac{\Delta}{\Lambda^2}}}{(-1-\beta)^5} p^4 \left(\frac{-2}{\Lambda^2}\right)^2  .
\end{aligned}
\end{eqnarray}
\section{}\label{sec:amplitudes}
\label{app:2}
The corresponding short distance processes can be represented as
\begin{eqnarray}
\label{FM}
&\mathcal{M}(B\rightarrow D^{(*)} M) &= \frac{G_F}{\sqrt{2}}V_{cb}V_{uq}^* ( a_{1} \bra{D^{(*)}}(\bar{c}b)_{V-A}\ket{B}\bra{M}(\bar{q}u)_{V-A}\ket{0} \notag\\
& &+a_{2} \bra{M}(\bar{q}b)_{V-A}\ket{B}\bra{D^{(*)}}(\bar{c}u)_{V-A}\ket{0}),\\
&\mathcal{M}(B\rightarrow M_1 M_2) &= \frac{G_F}{\sqrt{2}}V_{ub}V_{ud}^* ( a_{1} \bra{M_1}(\bar{u}b)_{V-A}\ket{B}\bra{M_2}(\bar{d}u)_{V-A}\ket{0} \notag\\
& &+a_{2} \bra{M_2}(\bar{d}b)_{V-A}\ket{B}\bra{M_1}(\bar{u}u)_{V-A}\ket{0})\\
&&+\frac{G_F}{\sqrt{2}}V_{tb}V_{td}^*\Big(a_3' \bra{M_1}(\bar{d}b)_{V-A}\ket{B}\bra{M_2}(\bar{q_i}q_i)_{V-A}\ket{0}\notag\\
&&+a_5' \bra{M_1}(\bar{d}b)_{V-A}\ket{B}\bra{M_2}(\bar{q_i}q_i)_{V+A}\ket{0}\Big),\notag\\
&\mathcal{M}(B\to \eta_c M)&=\frac{G_F}{\sqrt{2}}(V_{tb}V_{td}^*(a_3'-a_5' )-V_{cb}V_{cd}^*a_1 ) \langle\eta_c(p)|\overline{c}\gamma_\mu\gamma_5c|0\rangle \langle M(q)|\overline{d}\gamma_\mu b|B(p')\rangle, \\
&\mathcal{M}(B\to J/\psi M)&=\frac{G_F}{\sqrt{2}}(V_{cb}V_{cd}^*a_1-V_{tb}V_{td}^*(a_3'+a_5' ) )
\langle J/\psi(p)|\overline{c}\gamma_\mu c|0\rangle \langle M(q)|\overline{d}\gamma_\mu b|B(p')\rangle .
\end{eqnarray}
We write the amplitudes of Cabibbo-suppressed PP processes related to transitions $b\to c\bar us$, $b\to c\bar cd$, and $b\to u\bar u d$.
\begin{eqnarray}
\mathcal{M}(\overline B_s^0 \to \overline{D}^0T^0_{DK}/T^0_{D_s\pi})&=&\mathcal{M}_{2}^{DK}(J/\psi,K^0;D^0)_{V_{cd}}-\mathcal{M}^{DK}_{3,5}(J/\psi,K^0;D^0)_{V_{td}} \notag\\
&&+\mathcal{M}_{2}^{DK}(\rho^0/\omega,K^0;D^0)_{V_{ub}}-\mathcal{M}_{3,5}^{DK}(\rho^0/\omega,K^0;D^0)_{V_{td}}\notag\\
&&+\mathcal{M}_{1}^{D_s\pi}(D^{*-},D_s^+;\pi^-)_{V_{cd}}+\mathcal{M}_{1}^{D_s\pi}(K^{*+},\pi^-;D_s^+)_{V_{ub}},\notag\\
\mathcal{M}(B^- \to D_s^-T^0_{D_s\pi}/T^0_{DK})&=&\mathcal{M}_{2}^{D_s\pi}(J/\psi,\pi^-;D_s^+)_{V_{cd}}-\mathcal{M}^{D_s\pi}_{3,5}(J/\psi,\pi^-;D_s^+)_{V_{td}} \notag\\
&&+\mathcal{M}^{D_s\pi}_{1,2}(\rho^0/\omega,\pi^-;D_s^+)_{V_{ub}}-\mathcal{M}^{D_s\pi}_{3,5}(\rho^0/\omega,\pi^-;D_s^+)_{V_{td}}\notag\\
&&+\mathcal{M}^{DK}_{1}(D^{*-},D^0;K^0)_{V_{cd}},\notag\\
\mathcal{M}(B^+ \to D_s^-T^{++}_{D_s\pi}/T^{++}_{DK})&=&\mathcal{M}_{2}^{D_s\pi}(J/\psi,\pi^+;D_s^+)_{V_{cd}}-\mathcal{M}^{D_s\pi}_{3,5}(J/\psi,\pi^+;D_s^+)_{V_{td}}\notag\\
&&+\mathcal{M}^{D_s\pi}_{1,2}(\rho^0/\omega,\pi^+;D_s^+)_{V_{ub}}-\mathcal{M}^{D_s\pi}_{3,5}(\rho^0/\omega,\pi^+;D_s^+)_{V_{td}}\notag \\
&&+\mathcal{M}^{DK}_{1}(\overline{D}^{*0},D^+;K^+)_{V_{cd}},\notag\\
\mathcal{M}(B_s^0 \to \overline{D}^0T^{0}_{D\overline{K}}/T^{0}_{DK})&=&\mathcal{M}^{D\overline{K}}_{2}(J/\psi,\overline{K}^0;D^0)_{V_{cd}}-\mathcal{M}^{D\overline{K}}_{3,5}(J/\psi,\overline{K}^0;D^0)_{V_{td}}\notag \\
&&+\mathcal{M}^{D\overline{K}}_{2}(\rho^0/\omega,,\overline{K}^0;D^0)_{V_{ub}}-\mathcal{M}^{D\overline{K}}_{3,5}(\rho^0/\omega,,\overline{K}^0;D^0)_{V_{td}} \notag \\
&&+\mathcal{M}^{DK}_{1}(D_s^{*-},D^+;K^-)_{V_{cd}}+\mathcal{M}^{DK}_{1}(\rho^+,K^-;D^+)_{V_{ub}}, \notag \\
\mathcal{M}(\overline{B}_s^0 \to K^0T^{0}_{D\bar{K}}/T^{0}_{DK})&=&\mathcal{M}^{D\bar{K}}_{2}(\phi,D^0;\overline{K}^0)_{V_{us}}+\mathcal{M}^{DK}_{1}(D_s^{*+},K^-;D^+)_{V_{us}},\notag \\
\mathcal{M}(B^- \to \pi^-T^{0}_{D\bar{K}}/T^{0}_{DK})&=&\mathcal{M}^{D\bar{K}}_{1,2}(K^{*-},D^0;\overline{K}^0)_{V_{us}}+\mathcal{M}^{DK}_{1,2}(D^{*0},K^-;D^+)_{V_{us}},\notag \\
\mathcal{M}(\overline{B}^0\to \pi^0T^{0}_{D\overline{K}}/T^{0}_{DK})&=&\mathcal{M}_2^{D\overline{K}}(\overline{K}^{*0},D^0;\overline{K}^0)_{V_{us}}+\mathcal{M}_2^{D\overline{K}}(D^{*0},\overline{K}^{0};D^0)_{V_{us}}\notag\\
&&+\mathcal{M}_1^{DK}(K^{*-},D^+;K^-)_{V_{us}}+\mathcal{M}_1^{DK}(D^{*+},K^-;D^+)_{V_{us}},\notag\\
\mathcal{M}(\overline B_s^0 \to \overline K^0T^{0}_{DK}/T^{0}_{D_s\pi})&=&\mathcal{M}^{DK}_{2}(\phi,D^0;K^0)_{V_{us}}+\mathcal{M}^{D_s\pi}_{1}(K^{*-},D_s^+;\pi^-)_{V_{us}},
\end{eqnarray}
the amplitudes of Cabibbo-suppressed processes with VV configuration are
\begin{eqnarray}
&\mathcal{M}(\overline B_s^0 \to \overline{D}^0T^0_{D^{*}\overline{K}^{*}}/T^0_{D_s^{*}\rho})=&\mathcal{M}^{D^{*}\overline{K}^{*}}_{2}(\eta_c,\overline{K}^{*0};D^{*0})_{V_{cd}}-\mathcal{M}^{D^{*}\overline{K}^{*}}_{3,5}(\eta_c,\overline{K}^{*0};D^{*0})_{V_{td}}\notag\\
&&+\mathcal{M}^{D^{*}\overline{K}^{*}}_{2}(J/\psi,\overline{K}^{*0};D^{*0})_{V_{cd}}-\mathcal{M}^{D^{*}\overline{K}^{*}}_{3,5}(J/\psi,\overline{K}^{*0};D^{*0})_{V_{td}} \notag\\
&&+\mathcal{M}^{D_s^{*}\rho}_{1}(D^-,D_s^{*+};\rho^-)_{V_{cd}}+\mathcal{M}^{D_s^{*}\rho}_{1}(D^{*-},D_s^{*+};\rho^-)_{V_{cd}}, \notag\\
&\mathcal{M}(B^- \to D_s^-T^0_{D_s^{*}\rho}/T^0_{D^*K^*})=&\mathcal{M}^{D_s^{*}\rho}_{2}(\eta_c,\rho^-;D_s^{*+})_{V_{cd}}-\mathcal{M}^{D_s^{*}\rho}_{3,5}(\eta_c,\rho^-;D_s^{*+})_{V_{td}}\notag\\
&&+\mathcal{M}^{D_s^{*}\rho}_{2}(J/\psi,\rho^-;D_s^{*+})_{V_{cd}}-\mathcal{M}^{D_s^{*}\rho}_{3,5}(J/\psi,\rho^-;D_s^{*+})_{V_{td}} \notag\\
&&+\mathcal{M}^{D^*K^*}_{1}(D^-,D^{*0};K^{*0})_{V_{cd}}+\mathcal{M}^{D^*K^*}_{1}(D^{*-},D^{*0};K^{*0})_{V_{cd}},\notag\\
&\mathcal{M}(B^+ \to D_s^-T^{++}_{D_s^{*}\rho}/T^0_{D^*K^*})=&\mathcal{M}^{D_s^{*}\rho}_{2}(\eta_c,\rho^+;D_s^{*+})_{V_{cd}}-\mathcal{M}^{D_s^{*}\rho}_{3,5}(\eta_c,\rho^+;D_s^{*+})_{V_{td}} \notag\\
&&+\mathcal{M}^{D_s^{*}\rho}_{2}(J/\psi,\rho^+;D_s^{*+})_{V_{cd}}-\mathcal{M}^{D_s^{*}\rho}_{3,5}(J/\psi,\rho^+;D_s^{*+})_{V_{td}}\notag\\
&&+\mathcal{M}^{D^*K^*}_{1}(\overline{D}^{0},D^{*+};K^{*+})_{V_{cd}}+\mathcal{M}^{D^*K^*}_{1}(\overline{D}^{*0},D^{*+};K^{*+})_{V_{cd}},\notag\\
&\mathcal{M}(B_s^0 \to \overline{D}^0T^{0}_{D^{*}K^{*}}/T^0_{D^{*}K^*})=&\mathcal{M}^{D^{*}K^{*}}_{2}(\eta_c,K^{*0};D^{*0})_{V_{cd}}-\mathcal{M}^{D^{*}K^{*}}_{3,5}(\eta_c,K^{*0};D^{*0})_{V_{td}} \notag\\
&&+\mathcal{M}^{D^{*}K^{*}}_{2}(J/\psi,K^{*0};D^{*0})_{V_{cd}}-\mathcal{M}^{D^{*}K^{*}}_{3,5}(J/\psi,K^{*0};D^{*0})_{V_{td}}\notag\\
&&+\mathcal{M}^{D^{*}K^*}_{1}(D_s^-,D^{*+};K^{*-})_{V_{cd}}+\mathcal{M}^{D^{*}K^*}_{1}(D_s^{*-},D^{*+};K^{*-})_{V_{cd}}, \notag\\
&\mathcal{M}(\overline{B_s}^0 \to K^0T^{0}_{D^{*}\overline{K}^{*}}/T^{0}_{D^{*}K^{*}})=&\mathcal{M}^{D^{*}\overline{K}^{*}}_{2}(\eta,D^{*0};\overline{K}^{*0})_{V_{us}}+\mathcal{M}^{D^{*}\overline{K}^{*}}_{2}(\phi,D^{*0};\overline{K}^{*0})_{V_{us}}\notag \\
&&+\mathcal{M}^{D^{*}K^{*}}_{1}(D_s^{+},K^{*-};D^{*+})_{V_{us}}+\mathcal{M}^{D^{*}K^{*}}_{1}(D_s^{*+},K^{*-};D^{*+})_{V_{us}},\notag \\
&\mathcal{M}(B_s^0 \to \overline{D}^0T^{0}_{D^{*}K^{*}}/T^{0}_{D_s^{*}\rho})=&\mathcal{M}^{D^{*}K^{*}}_{2}(\pi^0/\eta,K^{*0};D^{*0})_{V_{ub}}-\mathcal{M}^{D^{*}K^{*}}_{3,5}(\pi^0/\eta,K^{*0};D^{*0})_{V_{td}} \notag \\
&&+\mathcal{M}^{D^{*}K^{*}}_{2}(\rho^0/\omega,K^{*0};D^{*0})_{V_{ub}}-\mathcal{M}^{D^{*}K^{*}}_{3,5}(\rho^0/\omega,K^{*0};D^{*0})_{V_{td}} \notag \\
&&+\mathcal{M}^{D_s^{*}\rho}_{1}(K^{+},\rho^-;D_s^{*+})_{V_{ub}}+\mathcal{M}^{D_s^{*}\rho}_{1}(K^{*+},\rho^-;D_s^{*+})_{V_{ub}}, \notag \\
&\mathcal{M}(B^- \to D_s^-T^{0}_{D_s^{*}\rho}/T^{0}_{D^{*}\overline{K}^{*}})=&\mathcal{M}^{D_s^{*}\rho}_{1,2}(\pi^0/\eta,\rho^-;D_s^{*+})_{V_{ub}}-\mathcal{M}^{D_s^{*}\rho}_{3,5}(\pi^0/\eta,\rho^-;D_s^{*+})_{V_{td}}\notag \\
&&+\mathcal{M}^{D_s^{*}\rho}_{1,2}(\rho^0/\omega,\rho^-;D_s^{*+})_{V_{ub}}-M^{D_s^{*}\rho}_{3,5}(\rho^0/\omega,\rho^-;D_s^{*+})_{V_{td}}\notag \\
&&+\mathcal{M}_2^{D^{*}\overline{K}^{*}}(\overline{K}^{0},D^{*0};\overline{K}^{*0})_{V_{us}}+\mathcal{M}_2^{D^{*}\overline{K}^{*}}(\overline{K}^{*0},D^{*0};\overline{K}^{*0})_{V_{us}}\notag\\
&&+\mathcal{M}_2^{D^{*}\overline{K}^{*}}(D^{0},\overline{K}^{*0};D^{*0})_{V_{us}}+\mathcal{M}_2^{D^{*}\overline{K}^{*}}(D^{*0},\overline{K}^{*0};D^{*0})_{V_{us}}\notag\\
&&+\mathcal{M}_1^{D^{*}K^{*}}(K^{-},D^{*+};K^{*-})_{V_{us}}+\mathcal{M}_1^{D^{*}K^{*}}(K^{*-},D^{*+};K^{*-})_{V_{us}}\notag\\
&&+\mathcal{M}_1^{D^{*}K^{*}}(D^{+},K^{*-};D^{*+})_{V_{us}}+\mathcal{M}_1^{D^{*}K^{*}}(D^{*+},K^{*-};D^{*+})_{V_{us}}.\notag\\
&\mathcal{M}(\overline B_s^0 \to \overline K^0T^{0}_{D^*K^*}/T^{0}_{D_s^*\rho})=&\mathcal{M}^{D^*K^*}_{2}(\eta,D^{*0};K^{*0})_{V_{cb}V_{us}}+\mathcal{M}^{D^*K^*}_{2}(\phi,D^{*0};K^{*0})_{V_{cb}V_{us}}\notag\\
&&+\mathcal{M}^{D_s^*\rho}_{1}(K^{-},D_s^{*+};\rho^-)_{V_{cb}V_{us}}+\mathcal{M}^{D_s^*\rho}_{1}(K^{*-},D_s^{*+};\rho^-)_{V_{cb}V_{us}},
\end{eqnarray}
Combing the short-distance with long-distance effects, the specific forms of the amplitudes of triangle diagrams in Fig.\ref{Fig:2} are deduced as,
\begin{eqnarray}
&\mathcal{M}_a=&\int \frac{d^4 k_3}{(2\pi)^4} \mathcal{F}(k_3^2) g_{\scalebox{0.5}{$\rho KK$}} g_{\scalebox{0.5}{$T_{cs(\bar{s})}DK$}} \left(a_1 m_\rho f_\rho A_{\mu}^{B\rightarrow D}+a_2 i f_{D}k_{1}^{\sigma} A_{\mu\sigma}^{B\rightarrow \rho} \right)\left(-g^{\mu\nu}+\frac{k_2^\mu k_2^\nu}{m_\rho ^2} \right)(k_{3}-p_{2 })_{\nu},  \notag\\
&\mathcal{M}_b=&\int \frac{d^4 k_3}{(2\pi)^4} \mathcal{F}(k_3^2) g_{\scalebox{0.5}{$D^* D_s K$}} g_{\scalebox{0.5}{$T_{cs(\bar{s})} \pi D_s$}} \left(a_1 i f_{\pi}k_{1}^{\sigma} A_{\mu\sigma}^{B\rightarrow D^*} + a_2 m_{D^*} f_{D^*} A_{\mu}^{B\rightarrow \pi} \right)\left( -g^{\mu \nu}+\frac{k_2^\mu k_2^\nu}{m_{D^*}^2} \right)(k_{3}-p_{2})_{\nu} ,\notag\\
&\mathcal{M}_c=& -\int \frac{d^4 k_3}{(2\pi)^4} \mathcal{F}(k_3^2) \left(   g_{\scalebox{0.5}{$\pi K K^*$}} g_{\scalebox{0.5}{$T_{cs(\overline{s})} D^* K^*$}}  \left(a_1 i f_{\pi}k_{2}^{\mu} A_{\mu\nu}^{B\rightarrow D^*}+ a_2 m_{D^*} f_{D^*} A_{\nu}^{B\rightarrow \pi}\right) \left(-g^{\nu\rho}+\frac{k_1^\nu k_1^\rho}{m_{D^*} ^2} \right)          \right.      \notag\\
&&\left(-g^{\gamma\sigma}+\frac{k_3^\gamma k_3^\sigma}{m_{K^*} ^2} \right) (k_{2\gamma}+p_{2\gamma})g_{\sigma \rho}-g_{\scalebox{0.5}{$\rho K K^*$}} g_{\scalebox{0.5}{$T_{cs(\overline{s})} D^* K^*$}}\left(a_1 m_\rho f_\rho A_{\mu\nu}^{B\rightarrow D^*}+a_2 m_{D^*}f_{D^*} A_{\mu\nu}^{B\rightarrow \rho} \right) \notag\\
&&\left.  \left(-g^{\mu\sigma}+\frac{k_2^\mu k_2^\sigma}{m_{\rho} ^2} \right)  \left(-g^{\nu\delta}+\frac{k_1^\nu k_1^\delta}{m_{D^*} ^2} \right) \varepsilon_{\rho \sigma \omega \theta} k_{2}^{\rho}k_3^{\omega}  \left(-g^{\theta\xi}+\frac{k_3^\theta k_3^\xi}{m_{K^*} ^2} \right)g_{\xi\delta} \right) , \notag\\
&\mathcal{M}_d=&-\int \frac{d^4 k_3}{(2\pi)^4}\mathcal{F}(k_3^2) \left(  g_{\scalebox{0.5}{$D D_s^* K$}} g_{\scalebox{0.5}{$T_{cs(\overline{s})} \rho D_s^*$}}\left(a_1 m_{\rho} f_{\rho} A_{\nu}^{B\rightarrow D} + a_2 i f_{D} k_{2}^{\mu} A_{\mu\nu}^{B\rightarrow \rho}\right) \left(-g^{\nu\rho}+\frac{k_1^\nu k_1^{\rho}}{m_\rho^2}\right)   \right. \notag\\
&&\left(-g^{\gamma\sigma}+\frac{k_3^\gamma k_3^\sigma}{m_{D_s^*}^2}\right)(k_{2\gamma}+p_{2\gamma})g_{\sigma\rho} -g_{\scalebox{0.5}{$D^*D_s^*K$}}g_{\scalebox{0.5}{$T_{cs(\overline{s})}\rho D_s^*$}}\left(a_1 m_\rho f_\rho A_{\mu\nu}^{B\rightarrow D^*}+a_2 m_{D^*}f_{D^*} A_{\mu\nu}^{B\rightarrow \rho} \right)\notag\\
&& \left. \left( -g^{\mu \sigma}+\frac{k_1^\mu k_1^\sigma}{m_{\rho}^2} \right)\left( -g^{\nu \delta}+\frac{k_2^\nu k_2^\delta}{m_{D^*}^2} \right) \varepsilon_{\rho \sigma \omega \theta} k_2^\rho k_3^\omega \left( -g^{\theta \xi}+\frac{k_3^\theta k_3^\xi}{m_{D_s^*}^2} \right)g_{\xi\delta} \right) , \notag\\
&A_{\mu}^{I\rightarrow X}=&F_1\left(q^2\right)\left(P_{I\mu}+P_{X\mu}-\frac{m_I^2-m_X^2}{q^2} q_{\mu}\right)+F_2\left(q^2\right)\left(\frac{m_I^2-m_X^2}{q^2} q_{\mu}\right),\notag \\
&A_{\mu\nu}^{I\rightarrow X^*}=&\left(\frac{2 i }{m_I+m_{X^*}} \varepsilon_{\mu \nu \alpha \beta}  P_{X}^{ \alpha} P_{I}^\beta  \right)  A_0\left(q^2\right)- \left( 2 m_{X^*}  \frac{ q_\nu}{q^2} q_{\mu} \right) A_1\left(q^2\right)-\left(m_I+m_{X^*}\right)\left(g_{\mu\nu}-\frac{q_{\nu}}{q^2} q_{\mu}\right)    \notag\\
&&A_2\left(q^2\right)+ \frac{q_\nu}{m_I+m_{X^*}}\left(P_{I\mu}+P_{X^*\mu}-\frac{m_I^2-m_{X^*}^2}{q^2} q_{\mu}\right) A_3\left(q^2\right).
\end{eqnarray}
Here, $I$ and $X$ denote the initial-state and final-state mesons respectively.  $p$, $k_3$ refer to the momenta of the corresponding particles.


\end{document}